\documentclass[10pt,twocolumn,letterpaper]{article}

\usepackage{cvpr}              
\usepackage{multirow}
\usepackage{multicol}
\usepackage{caption}
\usepackage{subcaption}
\usepackage{diagbox}
\DeclareRobustCommand{\rchi}{{\mathpalette\irchi\relax}}
\newcommand{\irchi}[2]{\raisebox{\depth}{$#1\chi$}}

\usepackage[dvipsnames]{xcolor}

\definecolor{cvprblue}{rgb}{0.21,0.49,0.74}
\usepackage[pagebackref,breaklinks,colorlinks,citecolor=cvprblue]{hyperref}

\def\paperID{12471} 
\def\confName{CVPR}
\def\confYear{2024}

\title{Robust Active Speaker Detection in Noisy Environments}

\author{Siva Sai Nagender Vasireddy, Chenxu Zhang, Xiaohu Guo, Yapeng Tian
\and
The University of Texas at Dallas, Richardson, TX, USA
\and
{\tt\small \{sivasainagender.vasireddy, chenxu.zhang, xguo, yapeng.tian\}@utdallas.edu}
}

\begin{document}
\maketitle
\begin{abstract}
       This paper addresses the issue of active speaker detection (ASD) in noisy environments and formulates a robust active speaker detection (rASD) problem. Existing ASD approaches leverage both audio and visual modalities, but non-speech sounds in the surrounding environment can negatively impact performance. To overcome this, we propose a novel framework that utilizes audio-visual speech separation as guidance to learn noise-free audio features. These features are then utilized in an ASD model, and both tasks are jointly optimized in an end-to-end framework. Our proposed framework mitigates residual noise and audio quality reduction issues that can occur in a naive cascaded two-stage framework that directly uses separated speech for ASD, and enables the two tasks to be optimized simultaneously.
To further enhance the robustness of the audio features and handle inherent speech noises, we propose a dynamic weighted loss approach to train the speech separator. We also collected a real-world noise audio dataset to facilitate investigations. Experiments demonstrate that non-speech audio noises significantly impact ASD models, and our proposed approach improves ASD performance in noisy environments. The framework is general and can be applied to different ASD approaches to improve their robustness. Our code, models, and data will be released.
\end{abstract}
\section{Introduction}

Active Speaker Detection (ASD) is the task of identifying the visible speakers in each frame of a video. It is an essential multimodal problem, and both facial dynamics and speech characteristics in sound provide strong cues. 
Recent advancements in the task have been made possible by the use of multimodal fusion of audio and visual modalities~\cite{roth2020ava,tao2021someone,min2022learning}. The availability of the AVA ActiveSpeaker dataset~\cite{roth2020ava}, a large-scale audio-visual dataset, has been instrumental in advancing research in this area.
Active speaker detection has numerous applications, including Video Conferencing~\cite{cutler2020multimodal,saravi2010real}, Smart Homes~\cite{busso2005smart}, Human-Computer Interaction~\cite{borde2004vviswa,martin2022multimodality}, and Bio-metrics~\cite{nagrani2018seeing,kinnunen2005real}. 

\begin{figure}[t]
\begin{center}
   \includegraphics[width=\linewidth]{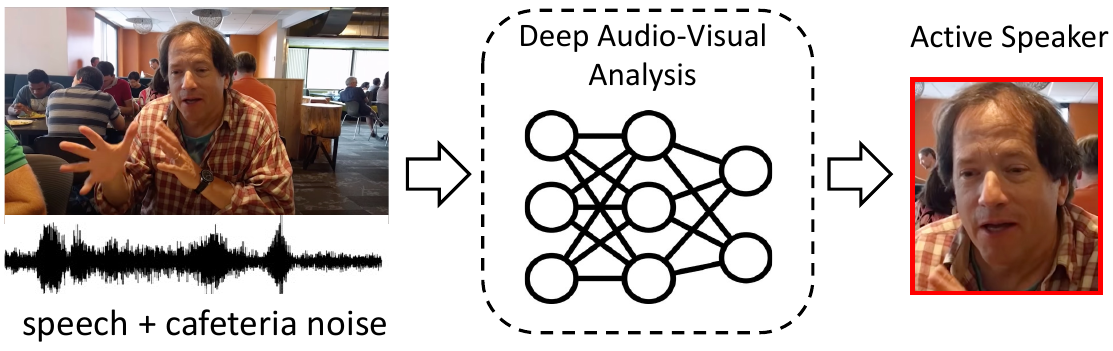}
\end{center}
\vspace{-6mm}
   \caption{ Given a video with both audio and visual tracks, we develop a robust deep audio-visual analysis model that can detect active speakers even in a noisy environment.} 
   \vspace{-7mm}
\label{fig:teaser}
\end{figure}

Existing approaches for active speaker detection focus on designing effective neural networks that can leverage temporal and multimodal context information in videos ~\cite{roth2020ava,tao2021someone,min2022learning,alcazar2020active,alcazar2022end}, in which both audio and visual modalities are fully exploited. 
However, the input modalities could be noisy and unreliable, particularly the audio. In real-world scenarios, non-speech sounds are common in the environment surrounding the active speaker. As shown in Fig.~\ref{fig:teaser}, the audio track contains both speech and strong cafeteria noise. Besides speech, current approaches will also encode undesired audio information into the representation, which can negatively impact active speaker detection performance.

In this paper, we formulate a robust active speaker detection (rASD) problem to address the issue.
The goal of the problem is to detect active speakers in videos with the presence of non-speech sounds in the surrounding environment. One possible solution to this problem is to separate the speech sound from the noisy audio mixture and then feed the separated sound to existing active speaker detection approaches. To train a separator that can perform such speech separation, we can use a mix-and-separate strategy~\cite{zhao2018sound,gao2019co} to generate training data. This involves randomly sampling non-speech audio and mixing it with clean speech sound. The separator can then be trained to separate speech from the mixture. Recent audio-visual sound separation and speech enhancement models~\cite{ephrat2018looking,zhao2018sound,gao2019co,afouras2018conversation,gao2021visualvoice} can be used as separators to address the problem.
However, speech separation and enhancement is a challenging task, and even state-of-the-art methods can leave behind residual noise in the separated speech. Additionally, the speech quality can be reduced compared to the original clean speech. Thus, this naive solution is not optimal.

Furthermore, the speech sounds in the training data used for active speaker detection~\cite{roth2020ava} can also be noisy as they are collected from web videos in the wild. Using these speech sounds as groundtruth for training the speech separation and enhancement models through the mix-and-separate strategy may lead to inferior performance.

To overcome these challenges, in this paper, we propose a novel framework for robust active speaker detection aimed at addressing the issue of audio noises in the surrounding environment of the active speaker. 
Instead of separating speech from noisy audio, we utilize audio-visual speech separation as guidance to learn noise-free audio features. These features are then utilized in an active speaker detection model. As a result, we can learn speech separation and active speaker detection simultaneously in a multi-task learning manner. Both tasks will use the same audio features, and the features will be enforced to be clean and helpful for the active speaker detection task. This approach mitigates the residual noise and audio quality reduction issues and enables the two tasks to be jointly optimized in an end-to-end framework. To handle inherent noise in speech sounds and further enhance the robustness of audio features, we propose a dynamic weighted loss approach to train the speech separator. In the approach, we reduce the importance of audio samples with inherent noise during training using weights that are dynamically generated.

To facilitate our investigations, we collected a real-world noise audio (RNA) dataset consisting of 1,350 non-speech sounds from 27 different categories. Experiments demonstrate that non-speech audio noises can significantly impact active speaker detection models. Our approach is capable of learning robust speech sound features, which can improve active speaker detection performance in noisy environments. Moreover, the proposed framework is general and can be applied to several different active speaker detection approaches to improve their robustness.

The main contributions of the work are three-fold:
\begin{itemize}
    \item  To the best of our knowledge, this is the first systematical study on robust active speaker detection considering the presence of real-world noise. We propose a novel framework for the problem that utilizes speech separation as guidance to learn noise-free speech sound features in a multi-task learning manner.  
    \item We propose a dynamic weighted loss to handle the inherent noise in speech sounds for learning more robust audio features for active speaker detection.

    \item We create a real-world noise audio dataset, named RNA, for our research. Extensive experiments on the large-scale AVA-ActiveSpeaker and RNA datasets can demonstrate the effectiveness and generalization capabilities of our proposed robust framework. 
\end{itemize}
\section{Related Work}

\textbf{Active Speaker Detection.} The pioneering work in active speaker detection is from Cutler and Davis~\cite{cutler2000look}. They use a time-delayed neural network to detect correlated audio-visual signals. Early follow-up works~\cite{everingham2009taking,saenko2005visual} depend only on visual modality and focus on facial gestures. However, ASD is fundamentally a multimodal task in which both audio and visual information are essential. With the availability of the large-scale AVA-ActiveSpeaker dataset~\cite{roth2020ava}, researchers have been able to develop effective audio-visual models for the ASD task. Large 3D convoluitional neural networks are used for audio-visual modeling in ~\cite{zhang2019multi}. To capture the temporal interactions between audio and visual features, non-local attention modules are explored~\cite{alcazar2020active}. TalkNet~\cite{tao2021someone} utilizes 3D networks and unimodal and cross-modal transformers achieving impressive results. Recently, high-performing audio and temporal visual models are exploited to improve ASD performance~\cite{kopuklu2021design}. Furthermore, a series of works~\cite{zhang2021unicon,min2022learning,leon2021maas,alcazar2022end, datta2022asd, liao2023light, xiong2022look, sharma2023audio} propose more advanced multimodal and temporal context modeling approaches. Diverging from prior research, we study the robustness of audio-visual speaker detection in the presence of non-speech noises and develop a general framework that can boost the robustness of existing ASD approaches.

\vspace{1mm}
\noindent
\textbf{Audio-Visual Sound Separation} Sound separation has been a long-standing problem in signal processing. In recent years, with the advent of deep learning-based joint audio-visual modeling, many deep audio-visual sound separation models have been developed. These models can separate visually indicated sounds from different sources, such as speech~\cite{ephrat2018looking,owens2018audio,gabbay2017visual,chung2020facefilter,gao2021visualvoice,afouras2018conversation}, musical instruments~\cite{zhao2018sound,xu2019recursive,gao2019co,gan2020music,tian2021cyclic}, and universal sound sources~\cite{gao2018learning,rouditchenko2019self,tzinis2021improving}. The mix-and-separate strategy~\cite{zhao2018sound,gao2019co} is commonly used to generate paired training samples for these models. These audio-visual sound separation methods leverage associations between audio and visual modalities, such as faces and speech, to separate individual sound sources from mixtures.
In this work, we explore how audio-visual speech separation can be used to improve active speaker detection under noisy conditions. Our framework generates noise-free speech features by transforming features from the separator. The two tasks are jointly optimized in a multi-task learning manner to facilitate robust audio feature learning.

\begin{figure*}[t]
\begin{center}
   \includegraphics[width=0.86\linewidth]{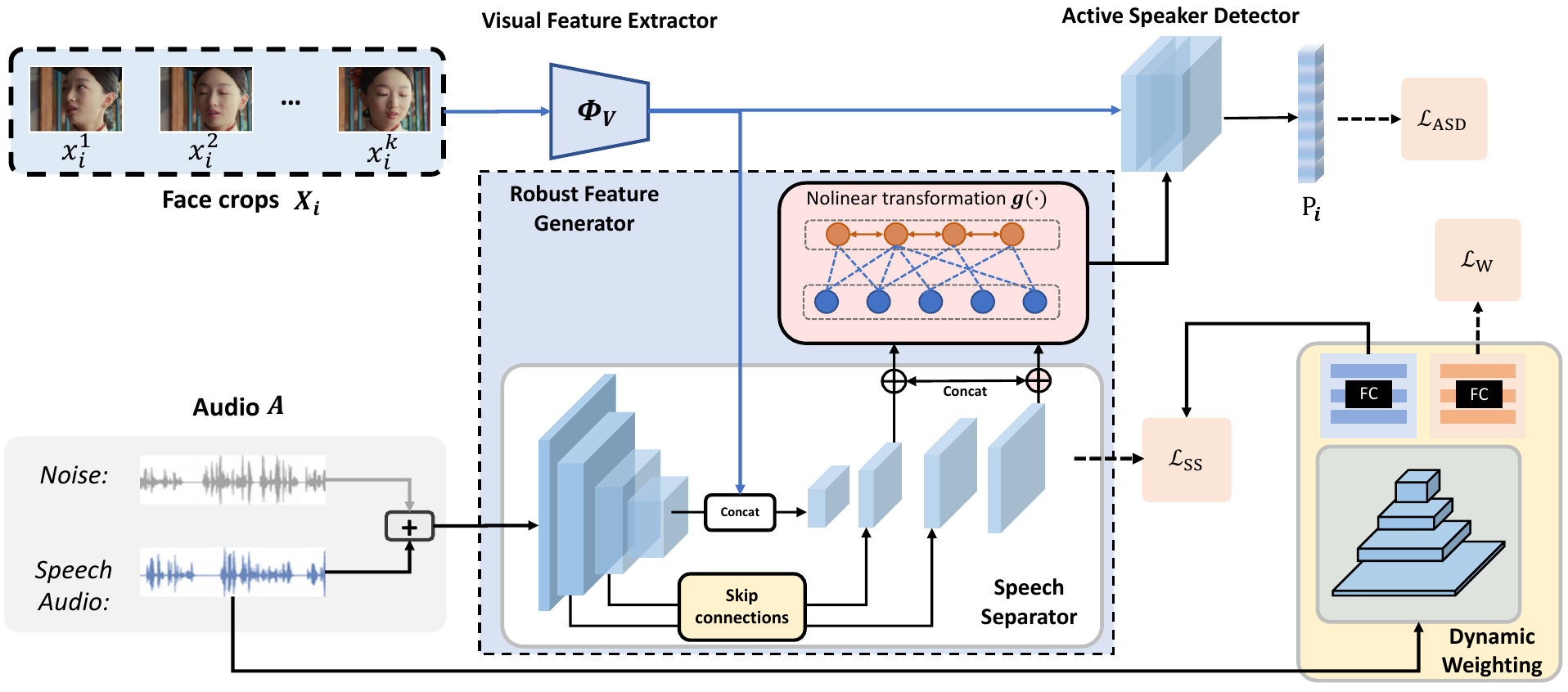}
\end{center}
\vspace{-5mm}
   \caption{
   The proposed robust active speaker detection framework. Upon the framework, we utilize an audio-visual speech separator to guide the learning of noise-free speech features for active speaker detection. The framework includes a nonlinear transformation $g(\cdot)$ to bridge the features between the separator and the detector. In addition, a dynamic weighting mechanism is employed to generate dynamic weights for the separation loss, which helps handle inherent speech noises. The framework is general and can be applied to improve the robustness of any existing audio-visual active speaker detectors.
   } 
   \vspace{-5mm}
\label{fig:framework}
\end{figure*}

\section{Method} \label{sec:methodoverview}

\subsection{Problem Formulation}  \label{sec:rasd}

\noindent
The goal of robust Active Speaker Detection (rASD)  is to identify visible speaking faces in a given video frame, while accounting for the presence of audio noise. Specifically, the audio clip corresponding to the video frame is a sound mixture: $A = A_{\texttt{speech}} + A_n$, which contains speech from active speakers and irrelevant audio noise from other sound sources in the surrounding environment. The task of rASD is highly challenging, as it requires effective utilization of the spatial, temporal, and multimodal contexts present in both audio and visual modalities while accounting for potential speech noises in the audio signal.

\subsection{Overview} \label{sec:proposedapproach}

\noindent
\textbf{Active Speaker Detection Pipelines. }
Most existing ASD methods~\cite{roth2020ava,tao2021someone,min2022learning,alcazar2020active,alcazar2021maas,alcazar2022end} can be generalized into four main modules: the visual feature extractor $\Phi_V$, the audio feature extractor $\Phi_A$, the audio-visual fusion module $\Phi_{AV}$, and the speaker detection module $\Phi_D$. Given a video with audio and visual tracks, these modules work together to detect the active speaker(s). Specifically, the inputs to these methods are a set of face crops $X_i = \{x_{i}^1, x_{i}^2, ..., x_{i}^k\}$ for each person $i$, with a sequence of $k$ face crops, and the corresponding audio $A$. The visual feature extractor $\Phi_V$ generates sets of feature representations $F_{v}^i = \Phi_V(X_i)$ of the face crops, while the audio feature extractor $\Phi_A$ generates sets of feature representations $F_{a}=\Phi_A(A)$ of their corresponding audio waveforms. The audio-visual fusion module $\Phi_{AV}$ then generates a set of audio-visual features $F_{av}^i=\Phi_{AV}(F_{a}, F_{v}^i)$ for each person $i$, which contains features corresponding to each of the face crops in $X_i$. Finally, the speaker detection module $\Phi_D$ generates the set of predictions $P_i=\Phi_D(F_{av}^i)$, which contains a prediction score $p_{i}^k$ for each of the face crops $x_{i}^k$. A loss function $\mathcal{L}_{ASD}$ is computed using the standard Cross-entropy loss.

\vspace{1mm}
\noindent
\textbf{Our Approach.} Rather than developing an advanced active speaker detection approach, our focus is on creating a sturdy audio-visual framework that can enhance the robustness of any existing active speaker detection method. Our approach centers on training a more robust feature generator to replace the existing $\Phi_A$. To improve the robustness of our model, we employ an audio-visual speech separation model to guide the robust audio feature learning. To address the presence of inherent audio noise in speech sounds and further reduce the effect of noise on encoded audio features, we introduce a dynamic weighted loss for separation. Figure~\ref{fig:framework} illustrates an overview of our rASD framework.

\subsection{Robust Audio Feature Generation} \label{sec:audiofeatureenhancement}

\noindent
We propose a robust audio feature generation module that produces robust audio features $F_a$ from noisy speech audio $A$. Our approach involves first computing the magnitude of the spectrogram $\rchi_A$ of the audio and then generating the robust audio features $F_a = \Phi_{RFG}(\rchi_A)$. The robust audio features $F_a$ can be integrated into active speaker detection pipelines to improve their robustness against real-world audio noise.
We incorporate a speech separator $\Phi_{SS}$ into our framework to ensure that the generated audio features contain noise-free speech information. The separator acts as guidance for our model.

\vspace{1mm}
\noindent
\textbf{Speech Separator. }Audio-visual sound separation~\cite{zhao2018sound,gao2019co,ephrat2018looking,afouras2018conversation,gao2021visualvoice} is an active area of research that aims to separate individual sound sources with the guidance of visible objects in a video. For speech, talking faces are typically the visible objects used. The Mix-and-Separate strategy is a commonly adopted approach to generate training data for the separation task. This strategy involves mixing a sound $A$, whose sound source is visible, with another randomly sampled audio to generate paired training data. During training, the audio mixture serves as the input to the separator, while the sound $A$ is used as the groundtruth. To learn a robust feature generator that can produce noise-free speech sound features for rASD, we use a speech separator $\Phi_{SS}$ as learning guidance, which can separate speech audio $S_\text{speech}$ from noisy audio $A$.

Similar to~\cite{zhao2018sound,gao2019co}, a U-Net~\cite{ronneberger2015u} architecture is used as a speech separator module $\Phi_{SS}$. We adopt the commonly used Mix-and-Separate strategy to train $\Phi_{SS}$. In our implementation, we generate noisy speech audio $A = A_{speech} + \alpha A_n$ by mixing speech audio $A_{speech}$ from the AVA-ActiveSpeaker~\cite{roth2020ava} dataset with randomly sampled real-world noise samples $A_n$, where $\alpha$ controls the noise level being added to the speech sound.
The separator takes the magnitude of the noisy spectrogram $\rchi_A  \in \mathbb{R}^{F \times T}$ as input and will output a separation mask $M_\text{pred} \in \mathbb{R}^{F \times T}$ with the help of tracked faces in video frames. 
 The spectrogram of the separated speech sound can be reconstructed by $\rchi_{A_{\text{speech}}} = M_\text{pred}\odot\rchi_A$, where $\odot$ represents element-wise multiplication operation. We compute the groundtruth ratio mask $M_{gt}$ as an element-wise ratio of $\rchi_{A_{speech}}$ and $\rchi_A$: $M_{gt}(p,q) = \frac{\rchi_{A_{\text{speech}}}(p,q)}{\rchi_A(p,q)}$. 
 To train the separator, we use an L1 loss function to compute the loss $L_{SS}$ between predicted and groundtruth masks:
\begin{equation} \label{eq:lss}
    \mathcal{L}_{SS} = \|M_{gt} - M_{pred}\|_1.
\end{equation}

\vspace{1mm}
\noindent
\textbf{Robust Feature Generator. }Our proposed approach to generating robust audio features is based on the premise that the set of intermediate feature maps $F_{in}$ from the decoder layers of the speech separator $\Phi_{SS}$ contains rich information about the speech sound $A_{\text{speech}}$. These feature maps $F_{in}$ carry the information required to separate the speech audio $A_{\text{speech}}$ from the input noisy audio $A$. This characteristic of $F_{in}$ makes it an ideal resource for generating speech audio features that are robust to the noise $A_n$. The robust audio features $F_a$ are generated from $F_{in}$ using transformation layers in robust feature generation module, $\Phi_{RFG}$.

The module $\Phi_{RFG}$ is built on top of the separator $\Phi_{SS}$ by adding a sequence of nonlinear transformations formulated by a function $g(\cdot)$ that generates robust audio features $F_a = g(F_{in})$ from a set of feature maps $F_{in}$ generated by $\Phi_{SS}$. $\Phi_{RFG}$ is trained in an end-to-end manner with the ASD loss.
Intuitively, the additional nonlinear transformations in $\Phi_{RFG}$ act as a bridge between the audio features that can separate speech audio from noise and the audio features that are required for the ASD task.

We investigated multiple combinations of feature maps from $F_{in}$ that can be utilized as input to the nonlinear transformation function, $g(\cdot)$, and found that a combination of feature maps, one closer to the bottleneck of the encoder-decoder U-Net architecture of $\Phi_{SS}$ and the other closer to the output of $\Phi_{SS}$ produces the best results. We hypothesize that this combination of feature maps represents the higher-level and lower-level features of the speech audio, where the feature map nearer to the separation output contains the cleanest speech sound information, while the one near the bottleneck contains more high-frequency speech patterns.

\subsection{Robustness to Inherent Noise}
As described in Sec. \ref{sec:audiofeatureenhancement}, the speech separator $\Phi_{SS}$ is trained with audio pairs $(A, A_{\text{speech}})$, with $A$ as input and $A_{\text{speech}}$ as target. However, audio in web videos could be noisy, and it may contain inherent noise $n_{in}$ even before adding external noise, \emph{i.e.}, $A_{\text{speech}}= A_c + A_{n_{in}}$, where $A_c$ is the clean speech component of $A_{\text{speech}}$. The presence of $A_{n_{in}}$ in $A_{\text{speech}}$ reduces the quality of speech representation in the feature maps $F_{in}$ of the speech separator $\Phi_{SS}$. The degradation in the quality of $F_{in}$ occurs because the feature maps also contain information about $A_{n_{in}}$ along with $A_c$ and groundtruth mask as the learning label becomes noisy.

To alleviate the negative impact of $A_{n_{in}}$ in our framework, we propose reducing the importance of audio samples with inherent noise by implementing a weighted separation loss approach. As different samples may have varying levels of noise, we present a dynamic weight generation approach to dynamically generate weights during training. This method helps to mitigate the adverse effects of inherent noise and enhances the robustness of our approach.

\vspace{1mm}
\noindent
\textbf{Weighted Loss For Separation.} 
A drawback of using the Mix-and-Separate strategy to separate speech from noise is the lack of availability of fully clean speech data. In web videos, a significant portion of speech sound contains either music or noise~\cite{47336}. The noisy labels will decrease speech separation performance and, consequently, affect the audio features which will be used for rASD.  To handle the inherent audio noises and strengthen the robustness of the speech sound representations, we use a weighted separation loss to optimize the speech separator.

In Eq.~\ref{eq:lss}, the separation loss is computed as the mean of losses of each sample in a batch, thus giving equal importance to each sample irrespective of the presence of inherent noise in speech audio. In our weighted loss approach, we assign a weight $w_k \in [0,1]$ to each training sample $k$:
\begin{equation}
    \label{eq:weightedloss}
   \mathcal{L}_{SS}^{k} = w_k \cdot \|{M}_{gt}^k - {M}_{pred}^k\|_1.
\end{equation}
The loss will be used to train the speech separator to increase the robustness of the features to inherent noise.

\vspace{1mm}
\noindent
\textbf{Dynamic Weight Generator.} One question that remained unanswered in implementing the weighted loss is how to obtain the weights. Sound types such as speech and non-speech can be easily obtained either through human annotations or an automatic audio tagging system~\cite{schmid2022efficient}. A naive strategy that we can employ is to set $w_k$ as a fixed scalar that is less than 1 for the samples with non-speech noise, thereby lowering the importance of these samples with inherent noises. However, samples can differ significantly, even when they all contain non-speech noise. 

Rather than using a fixed weight, we propose a dynamic weight generator, $\Phi_W$, that can be trained to predict weights based on the given audio input. This method enables us to generate weights that are tailored to the specific characteristics of each sample, further improving the robustness of our approach.

The architecture of $\Phi_W$ consists of a sequence of 2D convolutional layers, followed by two sets of fully connected layers, with ReLU being used as the activation function. It has two branches: one predicts the sound type class of $A$, while the other predicts the training sample weight. We train $\Phi_W$ using the loss $\mathcal{L}_W=\mathcal{L}_C + \frac{1}{b}\sum_{k=1}^b|w_k-1|$, where $b$ is the number of samples for which the loss is being computed, and $\mathcal{L_C}$ is the cross-entropy loss computed over the sound type classification. The second term serves to prevent an all-zero shortcut, as the weights are multiplied with the separation loss.

\subsection{Loss function for our approach to rASD}
\noindent
The final loss function of our robust ASD model is 
\begin{equation}
    \mathcal{L}=\lambda_1\mathcal{L}_{ASD} + \lambda_2\mathcal{L}_{SS}+\lambda_3\mathcal{L}_W,
\end{equation}
where $\lambda_1$, $\lambda_2$, and $\lambda_3$ are scalars to balance the loss terms, and they are empirically set as $0.1$, $1$, and $0.1$, respectively.

\section{Experiments}

\subsection{Datasets} \label{sec:datasets}
\noindent
To train and test rASD models, we utilize the AVA-ActiveSpeaker ~\cite{roth2020ava} as the ASD dataset and collect a Real-world Noise Audio dataset to simulate real-world noise. In addition, we show the efficacy of our approach in the real world using a subset of the AVA-ASD validation set comprising audio samples with inherent noise.

\vspace{1mm}
\noindent 
\textbf{AVA-ActiveSpeaker.}
The AVA-ActiveSpeaker~\cite{roth2020ava} is the first large-scale dataset annotated for the task of Active Speaker Detection. It includes 262 long Hollywood movies, with 120 movies in the training set, 33 in the validation set, and the remaining 109 in the testing set. The dataset has normalized bounding boxes for 5.3 million faces, all of which are manually curated from automatic detections. Facial detections are linked across time to produce face tracks that depict a single identity. Each face detection is labeled as speaking, speaking but not audible, or non-speaking.

Furthermore, the AVA-Speech~\cite{47336} contains sound type labels, including clean speech, speech with music, speech with noise, and no speech, to identify the inherent noise present in the audio of the AVA ActiveSpeaker dataset. In our experiments, we merge the labels speech with music and speech with noise into one label, speech with noise. These labels are used to train the dynamic weight generator. The distribution percentages of noise labels in the train and validation sets are provided in our appendix.

\vspace{1mm}
\noindent 
\textbf{RNA Dataset.} 
We collect a Real-world Noise Audio (RNA) dataset to simulate real-world noise in our experiments. We collect non-speech audio samples from the AudioSet~\cite{45857} dataset provided by Google. The AudioSet dataset provides labels of YouTube videos with annotations for 10-second clips, which are categorized into over 600 audio classes. Out of these, we selected 27 classes of audio which cover a broad range of real-world audio, such as, baby, dog, vacuum, indoor light machines, indoor running water, alarm clock, smoke alarm, dishes, cooking, horse, pig, goat, sheep, wind, water, fire, vehicle, emergency vehicle, aircraft, chainsaw, fireworks, guitar, violin, cello, flute, saxophone, and drums. For each selected audio class, we sample 50 audio labels. We further split the sample audio labels into the train, validation, and test sets, each of which consists of 1064, 133, and 133 clips, respectively. 

\vspace{1mm}
\noindent
\textbf{Real-world Data.}
We demonstrate the performance of our approach on datasets that represent real-world scenarios, without the addition of synthetic audio noise from the RNA dataset. To this end, we use a subset of the AVA-ActiveSpeaker validation set, named as AVA-Noise, which is collected by removing samples with clean speech audio from the complete validation set. The AVA-Noise dataset contains 7355 face tracks with a total of 0.6 million face bounding boxes extracted from 33 videos.

\vspace{1mm}
\noindent
\textbf{Training and Testing.} To incorporate real-world noise into the AVA-ActiveSpeaker dataset, we use noise audio from the RNA dataset. During the training stage, we randomly select a segment of a random noise audio $A_{n_{ij}}^t$ from the training set of the RNA dataset for the $i^{th}$ audio sample $A_{\text{speech}_i}^t$ from the AVA-ActiveSpeaker dataset. The corresponding noisy speech audio $A_{_{ij}}^t$ for the audio sample $A_{\text{speech}_i}^t$ in the $j^{th}$ epoch is generated as $A_{{ij}}^t = A_{\text{speech}_i}^t + \alpha_{ij}A_{n_{ij}}^t$, where $\alpha_{ij}\in[0,1]$ is a noise factor sampled from a uniform probability distribution over the range $[0,1]$. The noise factor $\alpha_{ij}$ controls the volume of noise added to the speech sample. This approach enables the neural network models to learn speech patterns in the presence of different levels of noise without overfitting to noise patterns.

During the testing phase, we first generate speech and noise pairs $(A_{speech_i}^v, A_{n_i}^v)$, where $A_{\text{speech}_i}^v$ is the $i^{th}$ speech audio sample from the validation set of the AVA ActiveSpeaker dataset, and $A_{n_i}^v$ is a randomly selected noise sample from the validation set of the RNA dataset. We generate six sets of noisy speech samples $A_{ij}^{v}$, such that $A_{ij}^v=A_{\text{speech}_i}^v + \alpha_j^vA_{n_i}^v$, where $j \in \{1, \dots, 6\}$ and $\alpha^v \in \{0, 0.2, 0.4, 0.6, 0.8, 1\}$. As the level of noise added to speech increases, the noisy speech validation set becomes more challenging. During real-world testing, we use data from AVA-Noise that contains inherent audio noises recorded in real-world scenarios.

\vspace{1mm}
\noindent 
\textbf{Evaluation.}
Following previous works on ASD~\cite{roth2020ava,tao2021someone,min2022learning}, we employ the official evaluation tool to compute mean average precision (mAP) as the metric for evaluation and report our results on the AVA-ActiveSpeaker validation set since the test set has not been publicly released.

\begin{figure*}[t]
\begin{center}
   \includegraphics[width=\linewidth]{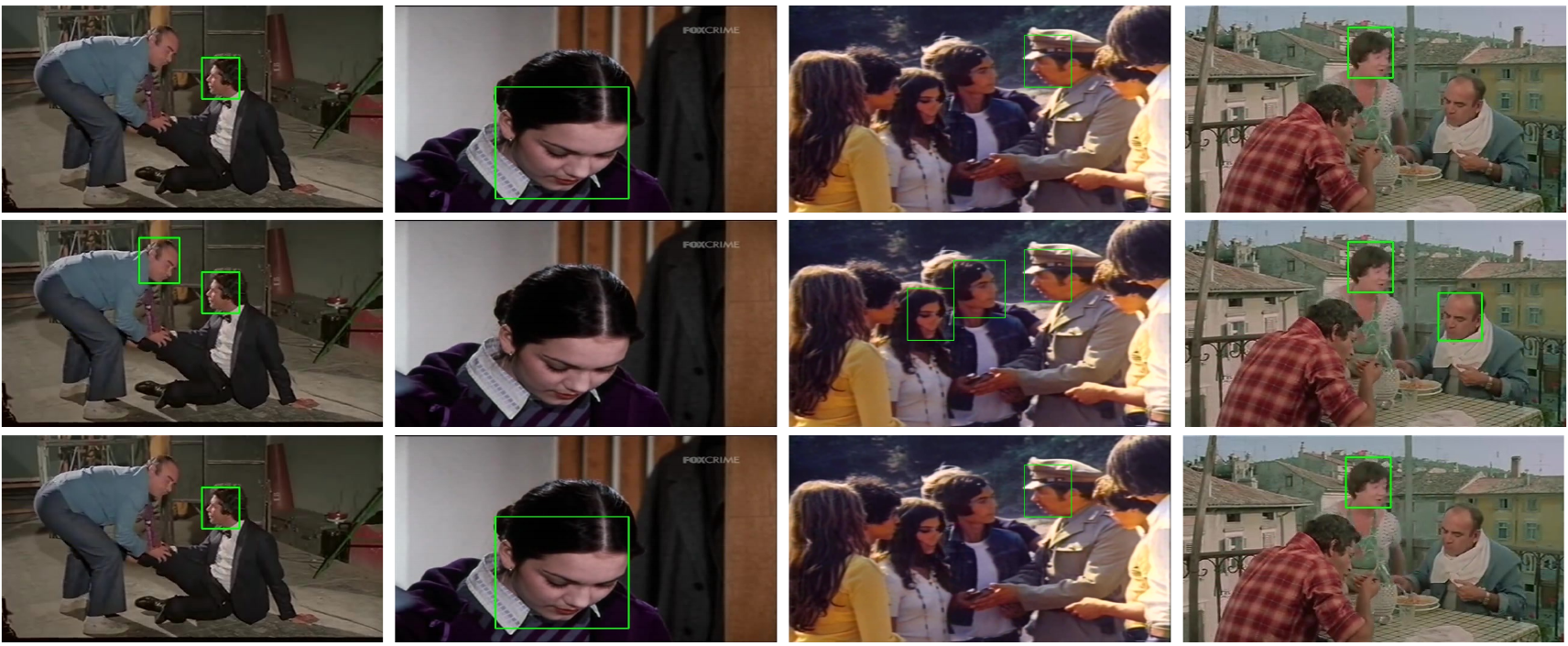}
\end{center}
\vspace{-5mm}
   \caption{Visual results of different methods under audio noises. From top to bottom, (1) GT: groundtruth active speakers; (2) Baseline: detected results by TalkNet with noisy training; (3) Ours: detected results by our framework with TalkNet. For the four examples, from left to right, four different non-speech sounds from aircraft, smoke alarms, pig, and emergency vehicle are added, respectively. Our framework can defend against audio noises and accurately detect active speakers } 
\label{fig:visual_res}

\end{figure*}

\subsection{Implementation Details} \label{sec:implementationdetails}

\noindent
In our experiments, we generate noisy speech audio $A$ from speech audio $A_{\text{speech}}$ and real-world noise audio $A_n$. We sample the input audio $A$ at 16 kHz, and our $\Phi_{RFG}$ module design allows the duration of audio to be variable. We compute the short-time Fourier transform (STFT) spectrogram from the input audio. The STFT spectrogram is computed with a window size of 1022 and a hop length of 160, resulting in a $512 \times T$ Time-Frequency spectrogram, where $T$ depends on the length of the input audio. We re-sample the spectrogram to $256 \times T$ for computational efficiency and construct the magnitude of the spectrogram $\rchi_{A} \in \mathbb{R}^{256\times T}$. We adopt the speech separator $\Phi_{SS}$ architecture from the sound separation network of ~\cite{gao2019co}.

For the weighted loss, $\mathcal{L}_{SS}$, the loss of each training sample is multiplied by a weighting factor which is based on the presence of inherent noises. The weight generation module $\Phi_W$ takes $\rchi_{A_{speech}} \in \mathbb{R}^{256 \times T}$ and generates classification probabilities to predict the audio type label of the class and a sample weight for the weighted loss function. 

We address the imbalance in the sound type classes by using a class-weighted cross-entropy loss where each class weight is inversely proportional to the number of labels of that class.
Our model is trained with Adam Optimizer~\cite{kingma2014adam} on an NVIDIA RTX A5000 GPU with a learning rate of 1e-4 which drops by a factor of 0.95 every epoch.

\begin{table*}
\begin{center}
\scalebox{0.86}{
\begin{tabular}{c|c|c c c|c c c|c c c|c c c|c c c}
\toprule
\multirow{2}{1em}{$\alpha$} &\multirow{2}{1em}{V} & \multicolumn{3}{c|}{MAAS~\cite{leon2021maas}} & \multicolumn{3}{c|}{ASC~\cite{alcazar2020active}} & \multicolumn{3}{c|}{SPELL~\cite{min2022learning}} & \multicolumn{3}{c|}{TalkNet~\cite{tao2021someone}} & \multicolumn{3}{c}{EASEE~\cite{alcazar2022end}} \\
                     &\multirow{9}{2em}{80.81}      & Orig & NT & Ours                & Orig & NT & Ours               & Orig & NT & Ours              & Orig & NT & Ours             & Orig & NT & Ours       \\
\midrule
0                    &                              & 84.75&84.11& \textbf{85.36}     &86.16 & 85.30 & \textbf{86.31}  & 90.78&91.07& \textbf{91.37}   & 92.31&92.25& \textbf{92.85}  & 90.26 & 89.81 &  \textbf{90.55} \\
0.2                  &                              & 77.86&79.81& \textbf{82.56}     &80.32 & 82.33& \textbf{84.38}   & 84.67&88.34& \textbf{89.71}   & 89.33&89.98& \textbf{91.37}  & 85.99 & 87.03 & \textbf{89.20}      \\
0.4                  &                              & 73.28&77.23& \textbf{80.59}     &76.15 & 80.24& \textbf{82.79}   & 80.49&86.54& \textbf{88.30}   & 87.50&88.51& \textbf{90.30}  & 83.26 & 84.65 & \textbf{87.71}   \\
0.6                  &                              & 69.97&75.38& \textbf{79.05}     &73.06 & 78.63& \textbf{81.54}   & 77.40&85.22& \textbf{87.19}   & 86.29&87.5& \textbf{89.39}   & 80.15 & 83.95 &  \textbf{86.46} \\
0.8                  &                              & 67.44&73.95& \textbf{77.44}     &70.65 & 77.33& \textbf{80.47}   & 75.01&84.21& \textbf{86.26}   & 85.37&86.67& \textbf{88.63}  & 79.44 & 83.01 &  \textbf{85.40} \\
1                    &                              & 65.42&72.71& \textbf{76.56}     &68.73 & 76.21& \textbf{79.51}   & 73.09&83.35& \textbf{85.39}   & 84.64&85.99& \textbf{87.97}  & 78.69 & 82.34 &  \textbf{84.98}   \\

\bottomrule
\multicolumn{2}{c|}{Average} &73.12 & 77.19& \textbf{80.26}& 75.85& 80.01& \textbf{82.50}& 80.24& 86.46& \textbf{88.04}& 87.57& 88.48& \textbf{90.09}& 82.97& 85.13& \textbf{87.38}\\
\bottomrule
\end{tabular}
}
\end{center}
\vspace{-4mm}
\caption{Results of robust active speaker detection using different approaches at different levels of $\alpha$. All of the models have been trained from scratch on the AVA-ActiveSpeaker dataset. Here, V represents a visual-only model adapted from TalkNet that is not affected by audio noises; Orig refers to the original model of each approach trained by following instructions on their respective GitHub pages; NT denotes noisy training. Our framework can be applied to each of the five different active speaker detection approaches, effectively enhancing their robustness and detection effectiveness. Compared with the models trained with noise in the training data (NT), our proposed approach obtains an improvement of 2.2\% mAP on average across all the baselines and all noise levels. The best results are highlighted.}
\label{tab:robustnesscomparison}
\vspace{-5mm}
\end{table*}

\begin{table*}
\begin{center}
\scalebox{0.86}{
\begin{tabular}{c|c c c|c c c|c c c|c c c|c c c}
\toprule
\multirow{2}{1em}{Data} & \multicolumn{3}{c|}{MAAS~\cite{leon2021maas}} & \multicolumn{3}{c|}{ASC~\cite{alcazar2020active}} & \multicolumn{3}{c|}{SPELL~\cite{min2022learning}} & \multicolumn{3}{c|}{TalkNet~\cite{tao2021someone}} & \multicolumn{3}{c}{EASEE~\cite{alcazar2022end}} \\
                     &   Orig & NT & Ours                & Orig & NT & Ours               & Orig & NT & Ours              & Orig & NT & Ours             & Orig & NT & Ours       \\
\midrule

AVA-Noise                  &  80.12 & 79.68 & \textbf{81.32}            & 82.25 & 81.73 & \textbf{83.21}          & 87.55 & 88.12 & \textbf{88.29}        & 88.33 & 88.50 & \textbf{89.36}        & 85.33 & 85.65 & \textbf{86.32} \\
\bottomrule
\end{tabular}
}
\end{center}
\vspace{-5mm}
\caption{ Detection results on AVA-Noise. 
We test model robustness in real-world scenarios without adding synthetic audio noise.}
\vspace{-5mm}
\label{tab:realworldrobustness}
\end{table*}

\subsection{Experimental Comparison}
In this section, we showcase the robustness of our approach by comparing its performance with other methods in the presence of audio noise. Additionally, we demonstrate that the learning capability of our approach is superior to that of the baseline methods. We present comparison results for both synthetic (Table~\ref{tab:robustnesscomparison}) and real-world (Table~\ref{tab:realworldrobustness}) scenarios.

\begin{table}
    \begin{center}
    \scalebox{0.58}{
       \begin{tabular}{ccccc|cccccc|c}
      \toprule
        \multirow{2}{1em}{SS} & \multirow{2}{4em}{Baseline} & \multirow{2}{2em}{RFG}  & \multirow{2}{2em}{DWL} & \multirow{2}{2em}{INL}  & \multirow{2}{1em}{0} & \multirow{2}{1em}{0.2} & \multirow{2}{1em}{0.4} & \multirow{2}{1em}{0.6} & \multirow{2}{1em}{0.8} & \multirow{2}{1em}{1} & \multirow{2}{2em}{Avg.} \\
        
             &                                      & & & & & & & & & \\
        \midrule
        &\checkmark  & & & &                               92.31 & 89.33 & 87.50 & 86.29 & 85.37 & 84.64 &87.57\\
                     \checkmark & \checkmark& & & &                        90.08 & 89.11 & 88.25 & 87.52 & 86.91 & 86.34 &88.04\\
                     & \checkmark &  \checkmark& & &             91.99 & 90.67 & 89.58 & 88.72 & 87.98 & 87.34&89.38\\
                    & \checkmark &  \checkmark& \checkmark& &     92.58 & 91.06 & 89.73 & 88.67 & 87.79 & 86.98 & 89.47 \\
                    
                    \checkmark & \checkmark & \checkmark & \checkmark & \checkmark & 92.48 & 90.87 & 89.68 & 88.77 & 88.02 & 87.36 & 89.53 \\
                    \bottomrule
                     & \checkmark &  \checkmark& \checkmark& \checkmark &  \textbf{92.85} & \textbf{91.37} & \textbf{90.30} & \textbf{89.39} & \textbf{88.63} & \textbf{87.97}&\textbf{90.09}\\
        \bottomrule
        \end{tabular}
        
    }
    \end{center}
    \vspace{-5mm}
    \caption{Ablation study on the proposed modules. Quantitative results of different models under different noise levels are shown. We use TalkNet~\cite{tao2021someone} as the baseline for this study. Here, SS denotes the use of separated speech as audio input, RFG refers to our robust feature generator, DWL represents our dynamic weight loss, and INL represents the inherent noise label classification task. The performance drop with SS demonstrates the adverse effects of using the speech generated from noisy audio by the speech separator. We can see that our full model (Baseline + RFG + DWL+ INL) achieves the best performance. The best results are highlighted. 
    }
    \label{tab:ablationstudy}
\end{table}

\vspace{1mm}
\noindent
\textbf{Baseline Models. }We select five recent ASD models, ASC~\cite{alcazar2020active}, MAAS~\cite{leon2021maas}, TalkNet~\cite{tao2021someone}, SPELL~\cite{min2022learning}, and EASEE~\cite{alcazar2022end} as our baselines. Two of them, TalkNet~\cite{tao2021someone} and EASEE~\cite{alcazar2022end} can be trained end-to-end, and the other three, MAAS~\cite{leon2021maas}, SPELL~\cite{min2022learning}, and ASC~\cite{alcazar2020active}, are trained in two stages, the feature representation stage and the context learning stage, respectively. With our comparison approach, we show that our proposed method is generalizable across multiple training strategies and can strengthen robustness on the ASD task irrespective of the training approach.

\vspace{1mm}
\noindent
\textbf{Comparison Approach. }For each of the baselines, we first train the method on the AVA-ActiveSpeaker dataset without adding any external noise to the training set and test the resulting models on the AVA ActiveSpeaker with noise added from the RNA dataset. In this step, we follow the instructions given on their respective GitHub repositories and present their performance (mAP) in the first column (Orig) of each method in Tab.~\ref{tab:robustnesscomparison}. 
We then add RNA noise to the training data of AVA-ActiveSpeaker (Sec ~\ref{sec:datasets}) and retrain the models. The performance of the models trained in the second step are in the second column (NT) of each of the methods.
In the final step, we integrate our robust audio feature generator into each of the baseline methods and train with noisy training data as in the previous step. The performance of the models from the final step are presented in the third column (Ours) of each of the methods. Through our experiments, we show that our approach is generalizable across multiple ASD methods and can enhance robustness in both synthetic and real-world noisy audio environments.

\begin{figure}[t]
\begin{center}
   \includegraphics[width=\linewidth]{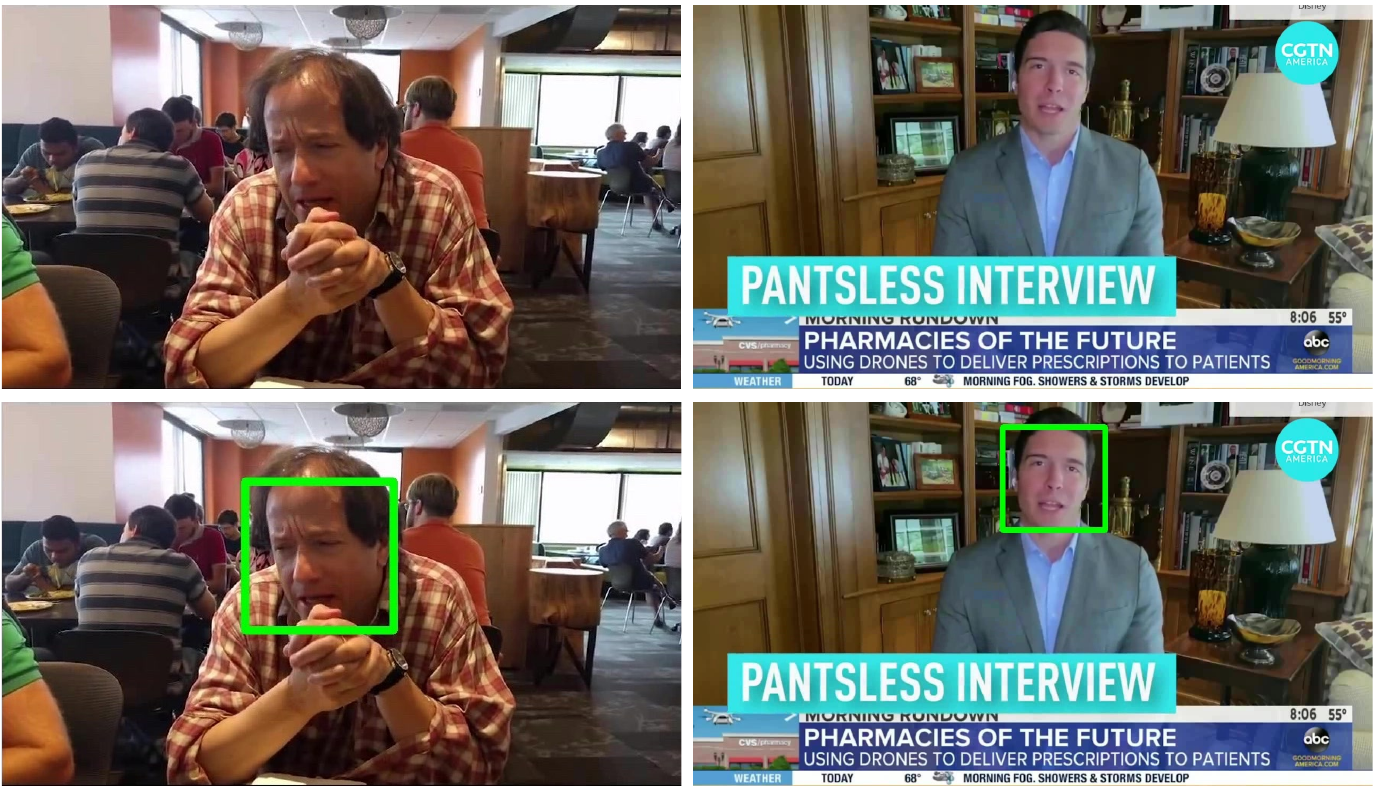}
\end{center}
\vspace{-5mm}
   \caption{Real-world visual examples, both of which are from video recordings without any additional non-speech sounds mixed in. The first example has strong cafeteria noise, while the second contains background music sounds. Our approach, TalkNet+Ours, can be applied to handle these real-world examples, as shown in the second row. In contrast, our baseline approach, TalkNet with noisy training, failed to perform well, as depicted in the first row.} 
\label{fig:visual_real}
\vspace{-7mm}
\end{figure}

\subsection{Results and Analysis}
\noindent
\textbf{Audio Noise Weakens ASD Performance.} 
As seen in Tab. ~\ref{tab:robustnesscomparison}, the performance of all five ASD approaches drops as the level of audio noise increases. When $\alpha = 1$, the performance of the original MAAS, ASC, SPELL, TalkNet, and EASEE models decreased by 19.3\%, 16.6\%, 17.4\%, 17.7\%, 7.7\%, and 11.6\%, respectively. Surprisingly, MAAS, ASC, SPELL, and EASEE as strong multimodal models, can achieve even worse performance than the visual-only unimodal baseline when $\alpha$ is large. These results demonstrate that non-speech sounds can significantly weaken ASD performance, and joint audio-visual modeling may not always be sufficient in a noisy environment.

\vspace{1mm}
\noindent
\textbf{Noisy Training.} The results show that NT models can generally improve ASD results once the input speech sound becomes noisy ($\alpha \geqslant 0.2$). This demonstrates that by adding randomly sampled audio noises into the training data, model robustness can be improved. Another observation is that NT models can decrease performance for most approaches when we do not add any external audio noises during testing. One possible reason for this is that NT models may overfit to some audio noises.

\vspace{1mm}
\noindent
\textbf{Our rASD framework is General and Effective.} Tab. ~\ref{tab:robustnesscomparison} demonstrates that our robust ASD framework can significantly improve the detection performance of different ASD approaches. Therefore, learning robust audio features is crucial for detecting active speakers in noisy environments.

We present additional visual results in Fig.~\ref{fig:visual_res}. With our robust framework, TalkNet can effectively detect active speakers in the presence of added audio noises (\eg, aircraft, smoke alarm, pig, and emergency vehicle sounds).

\vspace{1mm}
\noindent
\textbf{Real-world Data. }To further validate the effectiveness of our framework, we apply our model to real-world scenes. The quantitative results are shown in Tab.~\ref{tab:realworldrobustness}. Interestingly, the compared methods with noise training showed minor improvements or even performance drops. This is because simple noise training approaches cannot effectively suppress audio noises at different levels. The models were trained on more challenging noise sampled from the RNA dataset and tested on the relatively milder noise environment in the AVA-Noise dataset. In contrast, our approach can handle diverse real-world noisy scenes and outperforms the compared baselines. This demonstrates the robustness of our approach. Some visual results are shown in Fig.~\ref{fig:visual_real}. We can see that our model successfully detects talking faces in videos with cafeteria noises and background music, while the baseline model fails. These results further illustrate the effectiveness and generalization capability of our approach.

\vspace{1mm}
\noindent
\textbf{Baseline Replication. }Each baseline model (Orig column in Tab.~\ref{tab:robustnesscomparison}) was trained on the AVA-ActiveSpeaker dataset obtained from YouTube using the code from TalkNet~\cite{tao2021someone}. Discrepancies between our reported results and those in the original papers may be attributed to variations in training data due to unavailability of YouTube videos and modifications unrelated to the primary contributions of the original work. Nonetheless, we follow similar procedures in training the comparison models (Orig, NT, Ours) of each baseline.


\subsection{Ablation Study}
\noindent
Table~\ref{tab:ablationstudy} shows ablation study results on the proposed modules. Five different models are compared: (1) \textbf{Baseline}: We use the original TalkNet as a baseline; (2) \textbf{SS+Baseline}:  The separated speech sound is used as the audio input for TalkNet. This is a naive two-stage cascaded model that first performs audio-visual speech separation and then utilizes the separated speech sound as input for the active speaker detection model; (3) \textbf{Baseline+RFG}: It is our proposed model that uses audio-visual speech separation as guidance to learn noise-free speech sound features for the active speaker detection model; (4) \textbf{Baseline+RFG+DWL}: This model adopts dynamic weighted loss to handle inherent speech noises; (5) \textbf{Baseline+RFG+DWL+INL}: This is our full model that further adopts inherent noise label classification task as an auxiliary task to train the weight generator $\Phi_W$; (6) \textbf{SS+Baseline+RFG+DWL+INL}: This model shows the adverse effects of using speech separated from noise as input to our full model. From the table, we can see our full model achieves the best performance, which further demonstrates the effectiveness of our proposed framework.

\vspace{1mm}
\noindent
\textbf{Naive Cascaded Approach.} From Tab.~\ref{tab:ablationstudy}, we observe that {Baseline+SS} performs better than {Baseline} when stronger audio noises are added, but the overall improvement is relatively marginal. Additionally, when the noise levels are small ($\alpha = 0, 0.2$), this naive cascaded approach can even result in worse performance than {Baseline}. Our results indicate that utilizing separated speech can enhance the robustness of active speaker detection, particularly when strong noises exist. However, we also discovered that noise residuals could exist, and the speech quality may degrade in the separated speech, which limits its effectiveness.

\vspace{1mm}
\noindent
\textbf{Speech Separation as Guidance.} Compared to {Baseline+SS}, our proposed {Baseline+RFG} can improve active speaker detection performance across all different noise levels. Instead of using separated speech directly, this model utilizes speech separation as guidance to extract speech features for detection. The two tasks, namely separation and detection, are jointly optimized, and they can help each other to produce noise-free speech features that are suitable for detection and can reconstruct clean speech. As a result, the issues present in the {Baseline+SS} model are mitigated.

\vspace{1mm}
\noindent
\textbf{Dynamic Weighted Loss.} Additionally, our dynamic weighted loss approach is effective in handling inherent noises, which can further improve the robustness of our model (see last two rows in Tab. \ref{tab:ablationstudy}). It is worth noting that Baseline+RFG+DWL model can improve performance even without the need for additional noise labels. Further adding label guidance for the loss, our full model achieves the best performance. Due to its proficiency in handling noises in original speech sounds, the last two models in the table are the only noise-tolerant models that can improve active speaker detection performance when $\alpha = 0$.

\section{Conclusion}

In this paper, we present a systematic study of Active Speaker Detection (ASD) in the presence of audio noise and propose a robust Active Speaker Detection (rASD) framework. Our framework learns robust audio features using a speech separator as guidance and the dynamic weighted loss to capture individual speech sound characteristics. We believe that our proposed framework complements and boosts the state-of-the-art ASD methods rather than competing with them. To facilitate our investigations, we collected a new non-speech audio noise dataset, RNA. Our extensive experiments validate the superior robustness of our framework, which can significantly improve the accuracy of ASD methods in various real-world scenarios.

\clearpage
\setcounter{page}{1}
\maketitlesupplementary

\section*{Appendix}
In this appendix, we offer further clarification on our neural network architectures and present additional results. In Section~\ref{sec:noise}, we provide additional details about the audio noise labels in the AVA-ActiveSpeaker and AVA-Noise datasets.
In Section ~\ref{sec:arch}, we provide an in-depth description of our network architectures. Subsequently, we include supplementary ablation studies on the separated speech quality in Section~\ref{sec:ssq}, feature map selection for nonlinear transformation in Section~\ref{sec:nt}, and generated weights in Section~\ref{sec:weights}. Additionally, we present the first-stage results of two compared two-stage approaches in Section~\ref{sec:firststage}. Lastly, we discuss some limitations and future work in Section \ref{sec:limitations}.

\section{Audio Noise Labels}
\label{sec:noise}
\noindent
AVA-Speech~\cite{47336} contains sound type labels, including clean speech, speech with music, speech with noise, and no speech, to identify the inherent noise present in the audio of the AVA ActiveSpeaker dataset. In our experiments, we merge the labels speech with music and speech with noise into one label, speech with noise. These labels are used to train the dynamic weight generator. 
Table \ref{tab:noiselabeldistribution} shows the distribution percentages of noise labels in the AVA-ActiveSpeaker and the AVA-Noise datasets.

\begin{table}[tbh]
    \centering
    \begin{tabular}{l|ccc}
    \toprule
       Dataset (\#frames) &  Clean & With Noise & No Speech\\
    \midrule
        AVA-Train \hspace{0.005em} (2.7M)    & 17.70& 44.97 & 37.33\\
        AVA-Val \hspace{0.75em}  (0.8M)     & 21.33& 44.09& 34.58\\
        AVA-Full \hspace{0.45em}  (3.5M) & 18.51 & 44.77 & 36.72 \\
        AVA-Noise (0.6M)   & 0  & 56.04 & 43.96\\
    \bottomrule
    \end{tabular}
    \caption{Total duration percentages of each of the sound types in the AVA-ActiveSpeaker Dataset~\cite{roth2020ava} and the AVA-Noise datasets. The columns in the table correspond to the labels Clean Speech (Clean), Speech with Noise (With Noise), and No Speech (No Speech). The datasets represented in the rows of the table are the training set of AVA-ActiveSpeaker (AVA-Train), the validation set of AVA-ActiveSpeaker (AVA-Val), the full (train+val) AVA-ActiveSpeaker dataset used in our work (AVA-Full), and AVA-Noise dataset (AVA-Noise) which is a subset of the AVA-ActiveSpeaker validation set comprising of speech samples with inherent noise. Each visual frame in the dataset corresponds to a part of the audio with a noise label. The number of visual frames in each set of data is shown in parentheses in the first column.}
    \label{tab:noiselabeldistribution}
\end{table}

\section{Neural Network Architectures}
\label{sec:arch}
\noindent
In this section, we present the details of the neural network architecture details of our proposed modules. The dimensions provided here were used to generate robust features for the TalkNet~\cite{tao2021someone} ASD Model. Our code and pre-trained models will be released.
\begin{table}[t]
    \centering
    \begin{tabular}{c|c|c}
    \toprule
       Type/Stride  & Filter Shape & Input Size \\
    \midrule
       Conv1/s2  &    $4 \times 4 \times 1 \times 64$             & $1 \times 256 \times T$ \\
       Conv2/s2  &    $4 \times 4 \times 64 \times 128$             & $64 \times 128 \times T/2$ \\
       Conv3/s2  &    $4 \times 4 \times 128 \times 256$             & $128 \times 64 \times T/4$ \\
       Conv4/s2  &    $4 \times 4 \times 256 \times 512$             & $256 \times 32 \times T/8$ \\
       Conv5/s2  &    $4 \times 4 \times 512 \times 512$             & $512 \times 16 \times T/16$ \\
       Conv6/s2  &    $4 \times 4 \times 512 \times 512$             & $512 \times 8 \times T/32$ \\
       Conv7/s2  &    $4 \times 4 \times 512 \times 512$             & $512 \times 4 \times T/64$ \\
       Visual Concat &                                              & $512 \times 2 \times T/128$ \\
       DeConv1/s2 &   $4 \times 4 \times 1024 \times 512$             & $1024 \times 2 \times T/128$ \\
       DeConv2/s2 &   $4 \times 4 \times 1024 \times 512$             & $1024 \times 4 \times T/64$ \\
       DeConv3/s2 &   $4 \times 4 \times 1024 \times 512$             & $1024 \times 8 \times T/32$ \\
       DeConv4/s2 &   $4 \times 4 \times 1024 \times 256$             & $1024 \times 16 \times T/16$ \\
       DeConv5/s2 &   $4 \times 4 \times 512 \times 128$             & $512 \times 32 \times T/8$ \\
       DeConv6/s2 &   $4 \times 4 \times 256 \times 64$             & $256 \times 64 \times T/4$ \\
       DeConv7/s2 &   $4 \times 4 \times 128 \times 1$             & $128 \times 128 \times T/2$ \\
    \bottomrule
    \end{tabular}
    \caption{Detailed network architecture of the speech separator $\Phi_{SS}$. The input to the architecture is the magnitude of a spectrogram. The time dimension of the spectrogram can be of variable length. If $T < 256$, the time dimension is padded with zeros to make the time dimension at-least $256$. We use zero padding in the decoder to handle any size differences (due to variable T and the possibility of odd shape values of feature maps) during feature map concatenation. A visual feature of a face track (the average of all features of faces in the face track) is concatenated with the output of the encoder. The visual feature is a one-dimensional vector of size $512$ which is replicated spatially to match the shape of the bottleneck feature map.}

    \label{tab:speechseparatorarch}
\end{table}

\begin{table}
    \centering
    \begin{tabular}{c|c|c}
    \toprule
       Type/Stride  & Filter Shape & Input Size \\
    \midrule
       Conv1/S(2,1) & $3\times 3\times 640 \times 256$ & $640 \times 64 \times T/4$ \\
       Conv2/S(2,1) & $3\times 3\times 256 \times 128$ & $256 \times 32 \times T/4$ \\
       Conv3/S(2,1) & $3\times 3\times 128 \times 128$ & $128 \times 16 \times T/4$ \\
       AvgPool2D & $8\times 1$ & $128 \times 8 \times T/4$ \\
    \bottomrule
    \end{tabular}
    \caption{Detailed architecture of the nonlinear transformation $g(\cdot)$ used to generate robust audio features for the TalkNet ASD model. We concatenate feature maps from the separator and then use the concatenated feature maps as input for the nonlinear transformation layer to generate a 128D robust audio feature vector as output for ASD.  In the TalkNet ASD implementation, four audio frames are sampled from the input sound signal for each visual frame, and a single-dimension audio feature of size 128 is generated for each visual frame. The $g(\cdot)$ function used with the TalkNet ASD model generates a similarly shaped robust audio feature for each visual frame.}
    \label{tab:nonlineararchitecture}
\end{table}

\subsection{Robust Feature Generator}
\noindent
The robust feature generator $\Phi_{RFG}$ consists of two modules, the speech separator $\Phi_{SS}$ and the nonlinear transformation $g(\cdot)$. The speech separator is a UNet~\cite{ronneberger2015u} architecture with seven layers in the encoder and seven layers in the decoder as shown in Table \ref{tab:speechseparatorarch}. We use skip connections with feature map concatenations in our UNet implementation. The decoder portion of the speech separator $\Phi_{SS}$ generates six sets of feature maps $FM1, FM2, \dots, FM6,$ and a spectrogram mask as output, each of which is an output of the de-convolution (or up-convolution) layers DeConv1, DeConv2, \dots, DeConv6 and DeConv7 layers respectively. 

A concatenation of the feature maps, $FM3$ and $FM5$, is used as input to the nonlinear function $g(\cdot)$. The size of $FM3$ is increased to match the size of $FM5$ before concatenation by repeating its values locally. The architecture of $g(\cdot)$ that we used in robust feature generation for the TalkNet ASD model is shown in Table \ref{tab:nonlineararchitecture}. $g(\cdot)$, in this case, is designed to match the shape of generated audio features with the shape of the audio features used by TalkNet. Similarly, we modify the architecture of $g(\cdot)$ to generate features with the required shape for other ASD pipelines.

\begin{table}
    \centering
    \begin{tabular}{c|c|c|c}
    \toprule
       Index  & Inputs & Type/Stride & Input Size \\
    \midrule
       (1)  &  & Conv2D/S(2,1) & $1\times 256 \times T$\\
       (2)  & (1) & Conv2D/S(2,1) & $64\times 128 \times T$\\
       (3)  & (2) & Conv2D/S(2,1) & $64\times 64 \times T$\\
       (4)  & (3) & Conv2D/S(2,1) & $128\times 32 \times T$\\
       (5)  & (4) & Conv2D/S(2,1) & $128\times 16 \times T$\\
       (6)  & (5) & Conv2D/S(2,1) & $256\times 8 \times T$\\
       (7)  & (6) & Reshape(-1, T) & $256 \times 4  \times T$\\
       (8)  & \textbf{(7)} & Conv1D & $  1024 \times T$\\
       (9)  & (8) & Conv1D & $  256 \times T$\\
       (10)  & (9) & Conv1D & $  64 \times T$\\
       (11)  & (10) & Softmax & $  3 \times T$\\
       (12)  & \textbf{(7)} & Conv1D & $  1024 \times T$\\
       (13)  & (12) & Conv1D & $  256 \times T$\\
       (14)  & (13) & Conv1D & $  64 \times T$\\
       (15)  & (14) & Sigmoid & $  1 \times T$\\
    \bottomrule
    \end{tabular}
    \caption{Detailed architecture of the weight generator network. All of the 2D convolution kernels used in this architecture have the spatial dimension of ($3, 3$). The architecture takes a spectrogram with time dimension T and generates classification labels and weights for each of the audio frames. The architecture contains two fully connected heads, one of which produces the noise label classification scores (Layer 11) and the other produces weights (Layer 15). The fully connected heads are implemented using Conv1D layers, which enable us to generate weights and classification scores for all of the frames using only one neural network model.}
    \label{tab:weightgeneratorarchitecture}
\end{table}

\subsection{Dynamic Weight Generator}
\noindent
The dynamic weight generator $\Phi_W$ takes a spectrogram of audio (without any addition of audio from the RNA dataset (Section 4.1 of the main paper)) from the AVA Dataset and generates a weight for each audio frame\footnote[1]{A spectrogram consists of audio frames along its time dimension.} which is used in the Dynamic Weighted Loss (Section 3.4 of the main paper) along with the classification labels of the audio frame as described in Section 3.4 of the main paper. The architecture of our weight generator network is shown in Table \ref{tab:weightgeneratorarchitecture}.

\begin{table}
    \centering
    \begin{tabular}{c|c|c}
        \toprule
        Baseline & Diff. in \# Params & Avg. Inc. in mAP \\
        \midrule
        MAAS~\cite{leon2021maas} & $\approx$20M & 7.14\\
        ASC~\cite{alcazar2020active} & $\approx$20M & 6.65\\
        SPELL~\cite{min2022learning} & $\approx$20M & 7.8\\
        TalkNet~\cite{tao2021someone} & $\approx$8M & 2.52\\
        EASEE~\cite{alcazar2022end} & $\approx$20M & 4.41\\
        \midrule
        Avg. & 17.6M & 5.7 \\
        \bottomrule
    \end{tabular}
    \caption{Analysis of performance increment (average mAP over different noise levels) with number of parameters in the audio feature extractor. The table shows the additional number of parameters in the robust feature generator (RFG) compared to the audio feature generators in the baseline models in the second column and the average increase in performance over different noise levels in the third column. We can see an increase in performance of 1\% mAP in noisy environments for an additional 3M parameters on average.}
    \label{tab:numparams}
\end{table}

\subsection{Number of parameters} \label{sec:numparams}
In our experiments, we replace the audio feature extractor of the baseline model with our Robust Feature Generator (RFG). The entire RFG module has $\approx$32M parameters. In comparison, the audio feature extractor module: ResNet34 in TalkNet has $\approx$24M parameters and ResNet18 used in the other methods has $\approx$11M. $\approx$8M additional parameters are included compared to ResNet34-based audio feature extractor and $\approx$20M additional parameters are included compared to the ResNet18-based audio feature extractors. We noticed an increase in performance of 1\% mAP in noisy environments for an additional 3M parameters on average as shown in Table \ref{tab:numparams}.
\section{Ablation Study}

\begin{table}
    \centering
    \scalebox{0.85}{
    \begin{tabular}{c|ccc c c c }
    \toprule
         & 0 & 0.2 & 0.4 & 0.6 & 0.8 & 1 \\
        \midrule
         w/o DWL & 4.20 &1.92&1.60&1.46&1.37&1.31 \\
         w/ DWL  & \textbf{4.292} & 1.942 & 1.622& 1.469& 1.379& 1.325\\
         w/ DWL+INL & 4.287& \textbf{1.952}& \textbf{1.628} & \textbf{1.479} & \textbf{1.388} & \textbf{1.329}\\
         \bottomrule
    \end{tabular}
    }
    \caption{Speech quality comparison with PESQ score~\cite{miao_wang_2022_6549559}, without using the Dynamic Weighted Loss during training (w/o DWL), Dynamic Weighted Loss without instance noise label classification auxiliary task (w/ DWL), and Dynamic Weighted Loss with Instance Noise Label classification auxiliary task (w/ DWL+INL).  Higher PESQ scores indicate better speech quality. Our comparison examines the differences in speech quality at various noise levels, both with and without employing Dynamic Weighted Loss with Inherent Noise Label classification as an auxiliary task during the training process. The results demonstrate an increase in speech quality when incorporating Dynamic Weighted Loss, both with and without using Inherent Noise Label classification as an auxiliary task, which further substantiates the superiority of our robust audio features.}
    \label{tab:separatedspeechquality}
\end{table}

\subsection{Separated Speech Quality}
\label{sec:ssq}
\noindent
In Table~\ref{tab:separatedspeechquality}, we present sound separation results for models trained both with and without the proposed Dynamic Weighted Loss (DWL) with Inherent Noise Label (INL) classification as an auxiliary task during the training process. The findings reveal that incorporating DWL enhances the quality of separated speech sounds reconstructed from intermediate audio features. These results further underscore the superiority of the proposed DWL in boosting the robustness of audio features when dealing with inherent speech noise.

\begin{table*}[]
    \centering
    \begin{tabular}{c c c | c c c c c c | c}
        \toprule
        \multicolumn{3}{c|}{RFG} & \multirow{2}{1em}{0} & \multirow{2}{2em}{0.2} & \multirow{2}{2em}{0.4} & \multirow{2}{2em}{0.6} & \multirow{2}{2em}{0.8} & \multirow{2}{2em}{1} & \multirow{2}{2em}{Avg}\\
        FM3 & FM4 & FM5 & & & & & & & \\
        \midrule
        \checkmark & & &                          91.94 &                90.51  &        89.33           &      88.41             &           87.63         &        86.95        &        89.13                  \\
         & \checkmark & &                         91.67 &                90.10  &        89.02           &      88.16             &           87.44         &        86.78        &        88.86                  \\
         & & \checkmark &                         91.69 &                90.34  &        89.33           &      88.48             &           87.74         &        87.09        &        89.11                  \\
                   & \checkmark & \checkmark &    91.74 &                90.20  &        89.15           &      88.32             &           87.59         &        86.99        &        88.99                  \\
        \checkmark & \checkmark & \checkmark &    91.71 &                90.47  &        89.34           &      88.42             &           87.63         &        86.95        &        89.09                  \\
        \checkmark & & \checkmark &       \textbf{91.99}&        \textbf{90.67} &\textbf{89.58}          &\textbf{88.72}          &    \textbf{87.98}       &\textbf{87.34}       &\textbf{89.38}                  \\
        \bottomrule
    \end{tabular}
    \caption{Ablation study on feature map combination to use as input to the nonlinear function $g(\cdot)$. Robust active speaker detection performance on the validation set of the AVA ActiveSpeaker~\cite{roth2020ava} dataset combined with different levels of noise is shown. $FM3, FM4, $ and $FM5$ represent the third, fourth, and fifth feature maps of the decoder in the speech separator $\Phi_{SS}$ from its bottleneck. We can see that the combination of $FM3$ and $FM5$ achieves the best performance. The best results are highlighted.}
    \label{tab:featuremapcombinations}
\end{table*}
\subsection{Feature Map Selection for the nonlinear transformation $\boldsymbol{g(\cdot)}$}
\label{sec:nt}
\noindent
In this section, we justify our selection of feature maps from $F_{in}$ as described in Section 3.3 of the main paper. We investigated multiple combinations of feature maps from $F_{in}$ that can be utilized as input to the nonlinear transformation function $g(\cdot)$.  Out of the six feature maps, we have experimented with combinations of $FM3, FM4,$ and $FM5$, which are the third, fourth, and fifth feature maps, respectively, from the bottleneck of $\Phi_{SS}$. 

Using the original TalkNet ~\cite{tao2021someone} model with our robust feature generator ($\Phi_{RFG}$) to generate audio features, we present results of different combinations of feature maps in Table \ref{tab:featuremapcombinations}. When a selected combination contains multiple feature maps, the smaller feature maps are resized by replicating values locally to match the size of $FM5$, and the resulting feature maps are concatenated with $FM5$. In our work, we have selected the combination of feature maps $FM3$ and $FM5$ to form the input of the nonlinear function. 

\begin{table}
    \centering
    \scalebox{0.85}{
    \begin{tabular}{c|cccccc|c}
    \toprule
        \diagbox{$\alpha$}{w} & 0.6 & 0.7 & 0.8 & 0.85 & 0.9   & 1 &DW\\
        \midrule
        0                     &91.88&92.19&92.30&\textbf{92.38} & 92.23 & 91.99  &\textbf{92.85}  \\
        0.2                   &90.45&90.80&90.87&\textbf{90.92} & 90.81 & 90.67  &\textbf{91.37} \\
        0.4                   &89.38&89.64&89.71&\textbf{89.72} & 89.70 & 89.58  &\textbf{90.30} \\
        0.6                   &88.48&88.73&88.78&88.79 & \textbf{88.85} & 88.72   &\textbf{89.39}\\
        0.8                   &87.68&87.98&88.03&88.02 & \textbf{88.12} & 87.98   &\textbf{88.63}\\
        1                     &86.97&87.32&87.35&87.36 & \textbf{87.48} & 87.34   &\textbf{87.97}\\
    \bottomrule
    \end{tabular}
    }
    \caption{Results of training TalkNet~\cite{tao2021someone} with our robust feature generator with fixed weights for each noise type Note that the weight of speech with noise sample is w, the weight of clean speech or no speech sample is fixed with 1.  For reference the results of our model with dynamic weights (DW) are provided for reference. We can see that the best performance on each level of noise is distributed among higher weight values $\approx 1$, and our model with the dynamic weights achieves the best performance. Top-2 results for each level of noise are highlighted.}
    \label{tab:frameweightresults}
\end{table}

\begin{figure*}[tbh]
\begin{center}
   \begin{subfigure}[b]{0.45\textwidth}
   \centering
   \includegraphics[width=\textwidth]{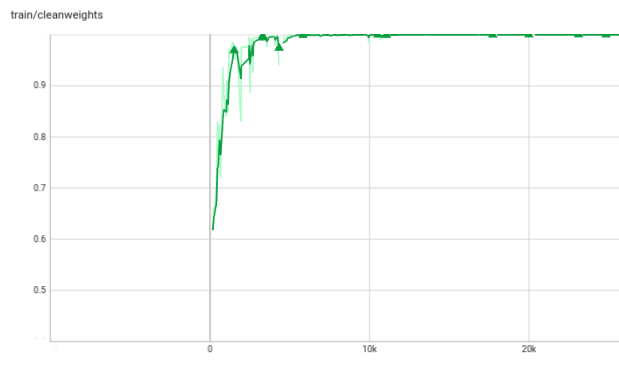}
   \caption{Mean clean training sample weight vs. number of training steps}
   \end{subfigure}
   \hfill
   \begin{subfigure}[b]{0.45\textwidth}
   \centering
   \includegraphics[width=\textwidth]{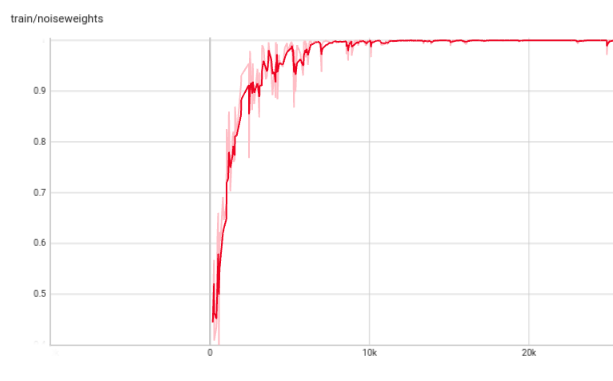}
   \caption{Mean noisy (inherent) training sample weight vs. number of training steps}
   \end{subfigure}
   \hfill
\end{center}
   \caption{Progression of mean training sample weight generated by $\Phi_W$ (Section 3.4 of the main paper) during training. The graphs clearly illustrate that the weights of clean speech samples are higher than the weights of noisy (inherent) speech samples during the initial epochs ($\approx 2000$ steps per epoch) and the weights converge to 1 in the later stages of training. This allows the speech separator $\Phi_{SS}$ to mainly learn from clean speech samples in the beginning and generalize to all speech samples in the later stages of training.}
\label{fig:weightgraphs}
\end{figure*}

\subsection{Weights generated by $\boldsymbol{\Phi_W}$}
\label{sec:weights}
\noindent
In this section, we present some results to support our claims about the effects of weighted loss in our approach. We have studied the effect of training sample weights in the loss function $\mathcal{L}_{SS}^k$ by fixing the weights of the samples based on the type of inherent noise. The results are presented in Table \ref{tab:frameweightresults}, where the weight w is assigned to samples with the label 'speech with noise' (Section 4.1 in the main paper). In Table \ref{tab:frameweightresults}, we can see that the best performance is distributed among higher values of w for different levels of noise. The observation also justifies the presence of the term $\frac{1}{b}\sum_{k=1}^b|w_k-1|$ in the loss function $\mathcal{L}_W$ (Section 3.4 of the main paper). This term encourages the weights of the training samples in the loss function $\mathcal{L}_{SS}^k$ to be $\approx 1$.

The progression of the weights generated by $\Phi_W$ during training is illustrated in Figure \ref{fig:weightgraphs}. We can see that the weights of clean speech training samples are higher than training samples labeled speech with noise (Section 4.1 in the main paper). This indicates that the parameters of the separator are trained with more importance to clean speech samples while generalizing to all of the training samples. 
\begin{table}
    \centering
    \begin{tabular}{c|ccc|ccc}
        \toprule
        \multirow{2}{1em}{$\alpha$} & \multicolumn{3}{c|}{ASC~\cite{alcazar2020active},MAAS~\cite{alcazar2021maas}} & \multicolumn{3}{c}{SPELL~\cite{min2022learning}}\\
                                    & Orig & NT & Ours &                                     Orig  &  NT & Ours\\
                                    \midrule
                                  0 & 79.51&78.59&\textbf{80.48} &                                    83.27       &  83.67   &   \textbf{85.14}   \\
                                 0.2& 73.16&74.52&\textbf{77.67} &                                     77.16      &  80.20   &   \textbf{82.76}   \\
                                 0.4& 69.24&72.19&\textbf{75.86} &                                     73.43      & 78.49    &   \textbf{81.16}   \\
                                 0.6& 66.54&70.66&\textbf{74.49} &                                     70.83      & 77.38    &   \textbf{79.99}   \\
                                 0.8& 64.52&69.51&\textbf{73.37} &                                     68.85      & 76.55    &   \textbf{79.09}   \\
                                  1 & 62.92&68.55&\textbf{72.4}  &                                     67.32      & 75.88    &   \textbf{78.32}   \\
                                  \bottomrule  
        
    \end{tabular}
    \caption{First stage results of ASC~\cite{alcazar2020active}, MAAS~\cite{alcazar2021maas}, and SPELL~\cite{min2022learning}. The results are reported on the validation set of AVA ActiveSpeaker combined with different levels of noise. \textbf{Orig} : The original method trained on AVA ActiveSpeaker. \textbf{NT} : The original method trained on AVA ActiveSpeaker with noise from the RNA dataset added to the original audio. \textbf{Ours} : Audio feature generator of the original method is replaced with our proposed robust feature generator and trained on AVA ActiveSpeaker with noise from RNA dataset added to original audio.}
    \label{tab:firststageresults}
\end{table}
\section{First Stage Results of Comparison Methods}
\label{sec:firststage}
\noindent
Three of our comparison methods, ASC~\cite{alcazar2020active}, MAAS~\cite{alcazar2021maas}, and SPELL~\cite{min2022learning}, are trained in two stages. In the first stage, they learn audio-visual representations of the inputs for the ASD task and they use spatial and temporal contexts to detect active speakers in the second stage. The final results of these methods are presented in Table 1 of the main paper. We present the first-stage results of these methods in this section, for completeness.

ASC~\cite{alcazar2020active} and MAAS~\cite{alcazar2021maas} use a two-stream network, following the work of AVA ActiveSPeaker~\cite{roth2020ava}, to generate audio-visual representations in the first stage. The first stage of SPELL~\cite{min2022learning} uses a similar feature representation, but they incorporate a TSM~\cite{lin2019tsm} module for computational efficiency. The first stage results of these methods are shown in Table \ref{tab:firststageresults}.

We have used ResNet-18 as the backbone of the two-stream network in the first stage of both of the methods. SPELL~\cite{min2022learning} reports an mAP of 88.0 in the first stage and an mAP of 94.2 after the second stage. The original GitHub repository provided by the authors does not contain the complete code to train their first-stage models. They have provided code to integrate TSM module into their first stage. The best results we could generate after following the steps provided by the authors are an mAP of 83.27 (Table \ref{tab:firststageresults}) in the first stage and an mAP of 90.78 (Table 1 in the main paper)
in the second stage. We used the results we could generate for our comparisons because we believe that the main contribution by the authors of SPELL~\cite{min2022learning} is their graph-based context-learning in the second stage and not the feature representation of the first stage. We believe that if we can replicate the first-stage results of SPELL~\cite{min2022learning}, then we can generate the second-stage results reported by the authors and show the robustness of our approach by generating test results similar to SPELL~\cite{min2022learning} in Table 1 of the main paper. 

\section{Limitations and Future Work} \label{sec:limitations}
In this work, we introduce a Robust Feature Generator(RFG) module and a novel training approach designed to enhance the robustness of the existing ASD models to address the rASD problem. We demonstrated the effectiveness of our RFG module experimentally (Table \ref{tab:robustnesscomparison}). 

The backbone of our RFG module is a U-Net module resulting in an increase in the number of parameters of the audio feature generator when compared to existing approaches that use ResNet18 and ResNet34 architectures as shown in Table \ref{tab:numparams}. In this work, we take the first step towards real-world ASD by designing the robust feature generator(RFG) and showing that robust features are crucial for real-world robustness. The next step in this direction is to explore parameter-efficient learning methods~\cite{houlsby2019parameter} to generate such robust features.

Our experimental results (Table \ref{tab:robustnesscomparison}) show significant improvement in performance at higher noise levels compared to the lower noise levels. Thus, there is scope for improved rASD approaches that also perform better in noise-free environments. In our future research, we will explore parameter-efficient robust feature generators that also show significant improvement in performance in noise-free environments.

{
    \small
    \bibliographystyle{ieeenat_fullname}
    \bibliography{rASD}

\begin{thebibliography}{46}
\providecommand{\natexlab}[1]{#1}
\providecommand{\url}[1]{\texttt{#1}}
\expandafter\ifx\csname urlstyle\endcsname\relax
  \providecommand{\doi}[1]{doi: #1}\else
  \providecommand{\doi}{doi: \begingroup \urlstyle{rm}\Url}\fi

\bibitem[Afouras et~al.(2018)Afouras, Chung, and Zisserman]{afouras2018conversation}
Triantafyllos Afouras, Joon~Son Chung, and Andrew Zisserman.
\newblock The conversation: Deep audio-visual speech enhancement.
\newblock \emph{arXiv preprint arXiv:1804.04121}, 2018.

\bibitem[Alc{\'a}zar et~al.(2020)Alc{\'a}zar, Caba, Mai, Perazzi, Lee, Arbel{\'a}ez, and Ghanem]{alcazar2020active}
Juan~Le{\'o}n Alc{\'a}zar, Fabian Caba, Long Mai, Federico Perazzi, Joon-Young Lee, Pablo Arbel{\'a}ez, and Bernard Ghanem.
\newblock Active speakers in context.
\newblock In \emph{Proceedings of the IEEE/CVF Conference on Computer Vision and Pattern Recognition}, pages 12465--12474, 2020.

\bibitem[Alc{\'a}zar et~al.(2021)Alc{\'a}zar, Caba, Thabet, and Ghanem]{alcazar2021maas}
Juan~Le{\'o}n Alc{\'a}zar, Fabian Caba, Ali~K Thabet, and Bernard Ghanem.
\newblock Maas: Multi-modal assignation for active speaker detection.
\newblock In \emph{Proceedings of the IEEE/CVF International Conference on Computer Vision}, pages 265--274, 2021.

\bibitem[Alc{\'a}zar et~al.(2022)Alc{\'a}zar, Cordes, Zhao, and Ghanem]{alcazar2022end}
Juan~Le{\'o}n Alc{\'a}zar, Moritz Cordes, Chen Zhao, and Bernard Ghanem.
\newblock End-to-end active speaker detection.
\newblock In \emph{European Conference on Computer Vision}, pages 126--143. Springer, 2022.

\bibitem[Borde et~al.(2004)Borde, Manza, Gawali, and Yannawar]{borde2004vviswa}
Prashant Borde, Ramesh Manza, Bharti Gawali, and Pravin Yannawar.
\newblock vviswa--a multilingual multi-pose audio visual database for robust human computer interaction.
\newblock \emph{International Journal of Computer Applications}, 137\penalty0 (4):\penalty0 25--31, 2004.

\bibitem[Busso et~al.(2005)Busso, Hernanz, Chu, Kwon, Lee, Georgiou, Cohen, and Narayanan]{busso2005smart}
Carlos Busso, Sergi Hernanz, Chi-Wei Chu, Soon-il Kwon, Sung Lee, Panayiotis~G Georgiou, Isaac Cohen, and Shrikanth Narayanan.
\newblock Smart room: Participant and speaker localization and identification.
\newblock In \emph{Proceedings.(ICASSP'05). IEEE International Conference on Acoustics, Speech, and Signal Processing, 2005.}, pages ii--1117. IEEE, 2005.

\bibitem[Chaudhuri et~al.(2018)Chaudhuri, Roth, Ellis, Gallagher, Kaver, Marvin, Pantofaru, Reale, Reid, Wilson, and Xi]{47336}
Sourish Chaudhuri, Joseph Roth, Dan Ellis, Andrew~C. Gallagher, Liat Kaver, Radhika Marvin, Caroline Pantofaru, Nathan~Christopher Reale, Loretta~Guarino Reid, Kevin Wilson, and Zhonghua Xi.
\newblock Ava-speech: A densely labeled dataset of speech activity in movies.
\newblock In \emph{Proceedings of Interspeech, 2018}, 2018.

\bibitem[Chung et~al.(2020)Chung, Choe, Chung, and Kang]{chung2020facefilter}
Soo-Whan Chung, Soyeon Choe, Joon~Son Chung, and Hong-Goo Kang.
\newblock Facefilter: Audio-visual speech separation using still images.
\newblock \emph{arXiv preprint arXiv:2005.07074}, 2020.

\bibitem[Cutler and Davis(2000)]{cutler2000look}
Ross Cutler and Larry Davis.
\newblock Look who's talking: Speaker detection using video and audio correlation.
\newblock In \emph{2000 IEEE International Conference on Multimedia and Expo. ICME2000. Proceedings. Latest Advances in the Fast Changing World of Multimedia (Cat. No. 00TH8532)}, pages 1589--1592. IEEE, 2000.

\bibitem[Cutler et~al.(2020)Cutler, Mehran, Johnson, Zhang, Kirk, Whyte, and Kowdle]{cutler2020multimodal}
Ross Cutler, Ramin Mehran, Sam Johnson, Cha Zhang, Adam Kirk, Oliver Whyte, and Adarsh Kowdle.
\newblock Multimodal active speaker detection and virtual cinematography for video conferencing.
\newblock In \emph{ICASSP 2020-2020 IEEE International Conference on Acoustics, Speech and Signal Processing (ICASSP)}, pages 4527--4531. IEEE, 2020.

\bibitem[Datta et~al.(2022)Datta, Etchart, Yadav, Hedau, Natarajan, and Chang]{datta2022asd}
Gourav Datta, Tyler Etchart, Vivek Yadav, Varsha Hedau, Pradeep Natarajan, and Shih-Fu Chang.
\newblock Asd-transformer: Efficient active speaker detection using self and multimodal transformers.
\newblock In \emph{ICASSP 2022-2022 IEEE International Conference on Acoustics, Speech and Signal Processing (ICASSP)}, pages 4568--4572. IEEE, 2022.

\bibitem[Ephrat et~al.(2018)Ephrat, Mosseri, Lang, Dekel, Wilson, Hassidim, Freeman, and Rubinstein]{ephrat2018looking}
Ariel Ephrat, Inbar Mosseri, Oran Lang, Tali Dekel, Kevin Wilson, Avinatan Hassidim, William~T Freeman, and Michael Rubinstein.
\newblock Looking to listen at the cocktail party: A speaker-independent audio-visual model for speech separation.
\newblock \emph{arXiv preprint arXiv:1804.03619}, 2018.

\bibitem[Everingham et~al.(2009)Everingham, Sivic, and Zisserman]{everingham2009taking}
Mark Everingham, Josef Sivic, and Andrew Zisserman.
\newblock Taking the bite out of automated naming of characters in tv video.
\newblock \emph{Image and Vision Computing}, 27\penalty0 (5):\penalty0 545--559, 2009.

\bibitem[Gabbay et~al.(2017)Gabbay, Shamir, and Peleg]{gabbay2017visual}
Aviv Gabbay, Asaph Shamir, and Shmuel Peleg.
\newblock Visual speech enhancement.
\newblock \emph{arXiv preprint arXiv:1711.08789}, 2017.

\bibitem[Gan et~al.(2020)Gan, Huang, Zhao, Tenenbaum, and Torralba]{gan2020music}
Chuang Gan, Deng Huang, Hang Zhao, Joshua~B Tenenbaum, and Antonio Torralba.
\newblock Music gesture for visual sound separation.
\newblock In \emph{Proceedings of the IEEE/CVF Conference on Computer Vision and Pattern Recognition}, pages 10478--10487, 2020.

\bibitem[Gao and Grauman(2019)]{gao2019co}
Ruohan Gao and Kristen Grauman.
\newblock Co-separating sounds of visual objects.
\newblock In \emph{Proceedings of the IEEE/CVF International Conference on Computer Vision}, pages 3879--3888, 2019.

\bibitem[Gao and Grauman(2021)]{gao2021visualvoice}
Ruohan Gao and Kristen Grauman.
\newblock Visualvoice: Audio-visual speech separation with cross-modal consistency.
\newblock In \emph{2021 IEEE/CVF Conference on Computer Vision and Pattern Recognition (CVPR)}, pages 15490--15500. IEEE, 2021.

\bibitem[Gao et~al.(2018)Gao, Feris, and Grauman]{gao2018learning}
Ruohan Gao, Rogerio Feris, and Kristen Grauman.
\newblock Learning to separate object sounds by watching unlabeled video.
\newblock In \emph{Proceedings of the European Conference on Computer Vision (ECCV)}, pages 35--53, 2018.

\bibitem[Gemmeke et~al.(2017)Gemmeke, Ellis, Freedman, Jansen, Lawrence, Moore, Plakal, and Ritter]{45857}
Jort~F. Gemmeke, Daniel P.~W. Ellis, Dylan Freedman, Aren Jansen, Wade Lawrence, R.~Channing Moore, Manoj Plakal, and Marvin Ritter.
\newblock Audio set: An ontology and human-labeled dataset for audio events.
\newblock In \emph{Proc. IEEE ICASSP 2017}, New Orleans, LA, 2017.

\bibitem[Houlsby et~al.(2019)Houlsby, Giurgiu, Jastrzebski, Morrone, De~Laroussilhe, Gesmundo, Attariyan, and Gelly]{houlsby2019parameter}
Neil Houlsby, Andrei Giurgiu, Stanislaw Jastrzebski, Bruna Morrone, Quentin De~Laroussilhe, Andrea Gesmundo, Mona Attariyan, and Sylvain Gelly.
\newblock Parameter-efficient transfer learning for nlp.
\newblock In \emph{International Conference on Machine Learning}, pages 2790--2799. PMLR, 2019.

\bibitem[Kingma and Ba(2014)]{kingma2014adam}
Diederik~P Kingma and Jimmy Ba.
\newblock Adam: A method for stochastic optimization.
\newblock \emph{arXiv preprint arXiv:1412.6980}, 2014.

\bibitem[Kinnunen et~al.(2005)Kinnunen, Karpov, and Franti]{kinnunen2005real}
Tomi Kinnunen, Evgeny Karpov, and Pasi Franti.
\newblock Real-time speaker identification and verification.
\newblock \emph{IEEE Transactions on Audio, Speech, and Language Processing}, 14\penalty0 (1):\penalty0 277--288, 2005.

\bibitem[K{\"o}p{\"u}kl{\"u} et~al.(2021)K{\"o}p{\"u}kl{\"u}, Taseska, and Rigoll]{kopuklu2021design}
Okan K{\"o}p{\"u}kl{\"u}, Maja Taseska, and Gerhard Rigoll.
\newblock How to design a three-stage architecture for audio-visual active speaker detection in the wild.
\newblock In \emph{Proceedings of the IEEE/CVF International Conference on Computer Vision}, pages 1193--1203, 2021.

\bibitem[Le{\'o}n-Alc{\'a}zar et~al.(2021)Le{\'o}n-Alc{\'a}zar, Heilbron, Thabet, and Ghanem]{leon2021maas}
Juan Le{\'o}n-Alc{\'a}zar, Fabian~Caba Heilbron, Ali Thabet, and Bernard Ghanem.
\newblock Maas: Multi-modal assignation for active speaker detection.
\newblock \emph{arXiv preprint arXiv:2101.03682}, 2021.

\bibitem[Liao et~al.(2023)Liao, Duan, Feng, Zhao, Yang, and Chen]{liao2023light}
Junhua Liao, Haihan Duan, Kanghui Feng, Wanbing Zhao, Yanbing Yang, and Liangyin Chen.
\newblock A light weight model for active speaker detection.
\newblock In \emph{Proceedings of the IEEE/CVF Conference on Computer Vision and Pattern Recognition}, pages 22932--22941, 2023.

\bibitem[Lin et~al.(2019)Lin, Gan, and Han]{lin2019tsm}
Ji Lin, Chuang Gan, and Song Han.
\newblock Tsm: Temporal shift module for efficient video understanding.
\newblock In \emph{Proceedings of the IEEE/CVF international conference on computer vision}, pages 7083--7093, 2019.

\bibitem[Martin et~al.(2022)Martin, Malpica, Gutierrez, Masia, and Serrano]{martin2022multimodality}
Daniel Martin, Sandra Malpica, Diego Gutierrez, Belen Masia, and Ana Serrano.
\newblock Multimodality in vr: A survey.
\newblock \emph{ACM Computing Surveys (CSUR)}, 54\penalty0 (10s):\penalty0 1--36, 2022.

\bibitem[Miao~Wang and ananda seelan(2022)]{miao_wang_2022_6549559}
Rafael G.~Dantas Miao~Wang, Christoph~Boeddeker and ananda seelan.
\newblock Pesq (perceptual evaluation of speech quality) wrapper for python users, 2022.

\bibitem[Min et~al.(2022)Min, Roy, Tripathi, Guha, and Majumdar]{min2022learning}
Kyle Min, Sourya Roy, Subarna Tripathi, Tanaya Guha, and Somdeb Majumdar.
\newblock Learning long-term spatial-temporal graphs for active speaker detection.
\newblock In \emph{Computer Vision--ECCV 2022: 17th European Conference, Tel Aviv, Israel, October 23--27, 2022, Proceedings, Part XXXV}, pages 371--387. Springer, 2022.

\bibitem[Nagrani et~al.(2018)Nagrani, Albanie, and Zisserman]{nagrani2018seeing}
Arsha Nagrani, Samuel Albanie, and Andrew Zisserman.
\newblock Seeing voices and hearing faces: Cross-modal biometric matching.
\newblock In \emph{Proceedings of the IEEE conference on computer vision and pattern recognition}, pages 8427--8436, 2018.

\bibitem[Owens and Efros(2018)]{owens2018audio}
Andrew Owens and Alexei~A Efros.
\newblock Audio-visual scene analysis with self-supervised multisensory features.
\newblock \emph{European Conference on Computer Vision (ECCV)}, 2018.

\bibitem[Ronneberger et~al.(2015)Ronneberger, Fischer, and Brox]{ronneberger2015u}
Olaf Ronneberger, Philipp Fischer, and Thomas Brox.
\newblock U-net: Convolutional networks for biomedical image segmentation.
\newblock In \emph{Medical Image Computing and Computer-Assisted Intervention--MICCAI 2015: 18th International Conference, Munich, Germany, October 5-9, 2015, Proceedings, Part III 18}, pages 234--241. Springer, 2015.

\bibitem[Roth et~al.(2020)Roth, Chaudhuri, Klejch, Marvin, Gallagher, Kaver, Ramaswamy, Stopczynski, Schmid, Xi, et~al.]{roth2020ava}
Joseph Roth, Sourish Chaudhuri, Ondrej Klejch, Radhika Marvin, Andrew Gallagher, Liat Kaver, Sharadh Ramaswamy, Arkadiusz Stopczynski, Cordelia Schmid, Zhonghua Xi, et~al.
\newblock Ava active speaker: An audio-visual dataset for active speaker detection.
\newblock In \emph{ICASSP 2020-2020 IEEE International Conference on Acoustics, Speech and Signal Processing (ICASSP)}, pages 4492--4496. IEEE, 2020.

\bibitem[Rouditchenko et~al.(2019)Rouditchenko, Zhao, Gan, McDermott, and Torralba]{rouditchenko2019self}
Andrew Rouditchenko, Hang Zhao, Chuang Gan, Josh McDermott, and Antonio Torralba.
\newblock Self-supervised segmentation and source separation on videos.
\newblock In \emph{Proceedings of the IEEE Conference on Computer Vision and Pattern Recognition Workshops}, pages 0--0, 2019.

\bibitem[Saenko et~al.(2005)Saenko, Livescu, Siracusa, Wilson, Glass, and Darrell]{saenko2005visual}
Kate Saenko, Karen Livescu, Michael Siracusa, Kevin Wilson, James Glass, and Trevor Darrell.
\newblock Visual speech recognition with loosely synchronized feature streams.
\newblock In \emph{Tenth IEEE International Conference on Computer Vision (ICCV'05) Volume 1}, pages 1424--1431. IEEE, 2005.

\bibitem[Saravi et~al.(2010)Saravi, Zafar, Edirisinghe, and Kalawsky]{saravi2010real}
Sara Saravi, Iffat Zafar, Eran~A Edirisinghe, and Roy~S Kalawsky.
\newblock Real-time speaker identification for video conferencing.
\newblock In \emph{Real-Time Image and Video Processing 2010}, pages 115--123. SPIE, 2010.

\bibitem[Schmid et~al.(2022)Schmid, Koutini, and Widmer]{schmid2022efficient}
Florian Schmid, Khaled Koutini, and Gerhard Widmer.
\newblock Efficient large-scale audio tagging via transformer-to-cnn knowledge distillation.
\newblock \emph{arXiv preprint arXiv:2211.04772}, 2022.

\bibitem[Sharma and Narayanan(2023)]{sharma2023audio}
Rahul Sharma and Shrkanth Narayanan.
\newblock Audio-visual activity guided cross-modal identity association for active speaker detection.
\newblock \emph{IEEE Open Journal of Signal Processing}, 2023.

\bibitem[Tao et~al.(2021)Tao, Pan, Das, Qian, Shou, and Li]{tao2021someone}
Ruijie Tao, Zexu Pan, Rohan~Kumar Das, Xinyuan Qian, Mike~Zheng Shou, and Haizhou Li.
\newblock Is someone speaking? exploring long-term temporal features for audio-visual active speaker detection.
\newblock In \emph{Proceedings of the 29th ACM International Conference on Multimedia}, pages 3927--3935, 2021.

\bibitem[Tian et~al.(2021)Tian, Hu, and Xu]{tian2021cyclic}
Yapeng Tian, Di Hu, and Chenliang Xu.
\newblock Cyclic co-learning of sounding object visual grounding and sound separation.
\newblock In \emph{Proceedings of the IEEE/CVF Conference on Computer Vision and Pattern Recognition}, pages 2745--2754, 2021.

\bibitem[Tzinis et~al.(2021)Tzinis, Wisdom, Remez, and Hershey]{tzinis2021improving}
Efthymios Tzinis, Scott Wisdom, Tal Remez, and John~R Hershey.
\newblock Improving on-screen sound separation for open-domain videos with audio-visual self-attention.
\newblock \emph{arXiv preprint arXiv:2106.09669}, 2021.

\bibitem[Xiong et~al.(2022)Xiong, Zhou, Zhang, Xie, Huang, and Zha]{xiong2022look}
Junwen Xiong, Yu Zhou, Peng Zhang, Lei Xie, Wei Huang, and Yufei Zha.
\newblock Look\&listen: Multi-modal correlation learning for active speaker detection and speech enhancement.
\newblock \emph{IEEE Transactions on Multimedia}, 2022.

\bibitem[Xu et~al.(2019)Xu, Dai, and Lin]{xu2019recursive}
Xudong Xu, Bo Dai, and Dahua Lin.
\newblock Recursive visual sound separation using minus-plus net.
\newblock In \emph{Proceedings of the IEEE International Conference on Computer Vision}, pages 882--891, 2019.

\bibitem[Zhang et~al.(2021)Zhang, Liang, Yang, Liu, Wu, Shan, and Chen]{zhang2021unicon}
Yuanhang Zhang, Susan Liang, Shuang Yang, Xiao Liu, Zhongqin Wu, Shiguang Shan, and Xilin Chen.
\newblock Unicon: Unified context network for robust active speaker detection.
\newblock In \emph{Proceedings of the 29th ACM International Conference on Multimedia}, pages 3964--3972, 2021.

\bibitem[Zhang et~al.(2019)Zhang, Xiao, Yang, and Shan]{zhang2019multi}
Yuan-Hang Zhang, Jingyun Xiao, Shuang Yang, and Shiguang Shan.
\newblock Multi-task learning for audio-visual active speaker detection.
\newblock \emph{The ActivityNet Large-Scale Activity Recognition Challenge}, 4, 2019.

\bibitem[Zhao et~al.(2018)Zhao, Gan, Rouditchenko, Vondrick, McDermott, and Torralba]{zhao2018sound}
Hang Zhao, Chuang Gan, Andrew Rouditchenko, Carl Vondrick, Josh McDermott, and Antonio Torralba.
\newblock The sound of pixels.
\newblock In \emph{Proceedings of the European conference on computer vision (ECCV)}, pages 570--586, 2018.

\end{thebibliography}
}

\end{document}